%% file: templateArxiv.tex
\documentclass[11pt]{article}

\usepackage[a4paper]{geometry}
\usepackage{bbm}
\usepackage{subcaption}

\usepackage{amssymb}
\usepackage{fix-cm}
\usepackage{authblk}
\usepackage{booktabs} 
\usepackage{indentfirst}
\usepackage{times}
\usepackage{graphicx}
\graphicspath{{Figs/}}
\usepackage{comment}
\usepackage{caption}
\usepackage[utf8]{inputenc} 
\usepackage[T1]{fontenc}  
\usepackage{url}   
\usepackage{booktabs}       
\usepackage{nicefrac}     
\usepackage{microtype}     
\usepackage{makecell}
\usepackage{xcolor}
\usepackage[framemethod=tikz]{mdframed}
\usepackage{mathtools}
\usepackage{amsmath,amsfonts,amssymb,amscd,xspace}
\usepackage{mathrsfs}
\usepackage{amsthm}
\usepackage[hyperfootnotes=false]{hyperref}
\usepackage{cleveref}
\usepackage{extarrows}
\usepackage{appendix}
\usepackage{tocbasic}
\crefformat{equation}{\textit{Eq.}(#2#1#3)}

\usepackage[shortlabels]{enumitem}
\usepackage{bbm}
\usepackage{bm}
\usepackage{booktabs}
\usepackage[ruled]{algorithm2e}
\usepackage{algpseudocode}

\usepackage{algorithm}


\graphicspath{{Figs/}}

\usepackage{graphicx}
\providecommand{\keywords}[1]
{
  \textbf{\textit{Keywords---}} #1
}

\newtheorem{theorem}{Theorem}
\newtheorem{lemma}{Lemma}
\newtheorem{corollary}{Corollary}

\theoremstyle{definition}
\newtheorem{definition}{Definition}
  
\title{\textbf{The Exact Boundary of Linear Optimality in Witsenhausen's Counterexample}}

\author{Yikai Lu and Tongxin Li
\thanks{Both authors are with the School of Data Science, The Chinese University of Hong Kong, Shenzhen, Guangdong, China
(\texttt{yikailu@link.cuhk.edu.cn},
\texttt{litongxin@cuhk.edu.cn}).}}

\begin{document}
\date{}
\maketitle

\begin{abstract}

    Witsenhausen’s counterexample has resisted a closed form solution for decades. However, it has remained unknown whether a mathematically tractable representation exists or not. In this paper, we solve this open problem, showing that the optimal policies are either linear or not tame and characterizing the exact regions where the two properties hold. Moreover, in the linear optimal region, we give an essentially unique closed form solution for the optimal policy, and show that no optimal first-stage policy is definable in any $o$-minimal expansion of the ordered real field in the nonlinear optimal region.

\end{abstract}


\section{Introduction}

\label{sec:introduction}

In $1968$, Witsenhausen~\cite{counterexample} constructed a linear quadratic Gaussian (LQG) team problem to show nonlinear policies can outperform linear policies in decentralized stochastic control with linear dynamics, quadratic costs, and Gaussian disturbances. This counterexample highlights the importance of information structure for linear optimality. It also reveals the signaling role of control actions, which shape both the system state and the information available to the second controller~\cite{MitterSahai1999,GroverParkSahai2013}.

\begin{figure}[ht]
    \centering
    \includegraphics[width = 0.6 \linewidth]{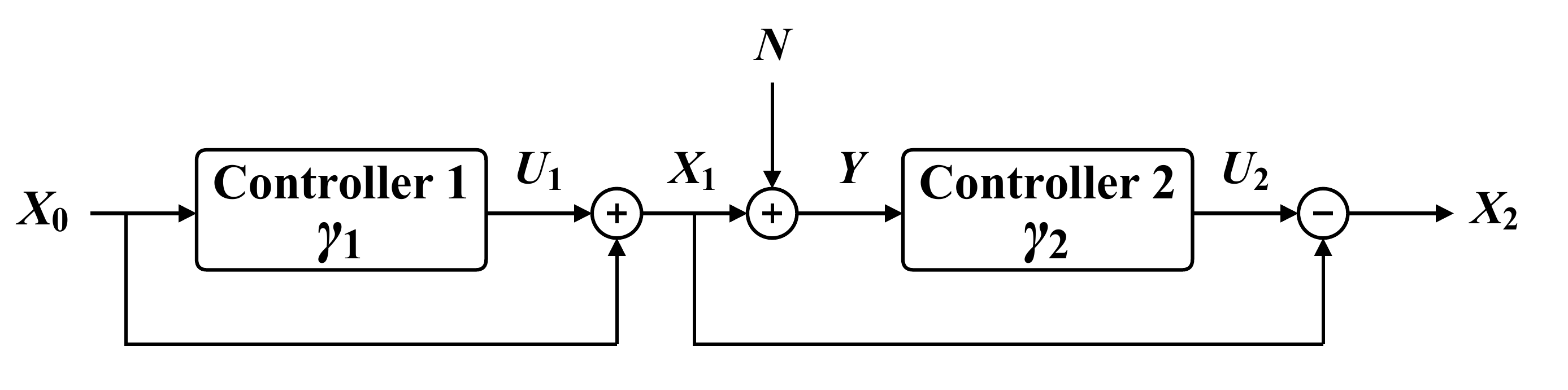}
    \caption{Problem statement of Witsenhausen's counterexample.}
    \label{fig:statement}
\end{figure}

In this paper, we identify an exact phase transition between linear and nonlinear optimal control in the classical scalar Gaussian problem. As summarized in Figure~\ref{fig:statement}, the problem concerns an initial state $X_0 \sim \mathcal N(0, \sigma ^2)$ with standard deviation $\sigma > 0$ and an independent observation noise $N \sim \mathcal N(0, 1)$. The first controller observes state $X_0$ and applies the control input $U_1 = \gamma_1(X_0),$ resulting in the immediate state $X_1 = X_0 + U_1$. The second controller observes noisy state $Y = X_1 + N$ and applies $U_2 = \gamma_2(Y)$, yielding the terminal state $X_2 = X_1 - U_2$. We write $f(x) = \gamma_1(x) + x$ and $g(x) = \gamma_2(x)$ for simplicity. For a positive first-stage control weight parameter $k > 0$, we seek Borel measurable policies $f$ and $g$ that minimizes
\begin{equation}
\label{eq:cost}
J_{k, \sigma}(f, g) = k ^2\mathbb E [(f  (X_0 ) - X_0) ^2 ] + \mathbb E [(X_1 - g(Y) ) ^2].
\end{equation}

\begin{figure}[t]
    \centering
    \resizebox{0.94\linewidth}{!}{\input{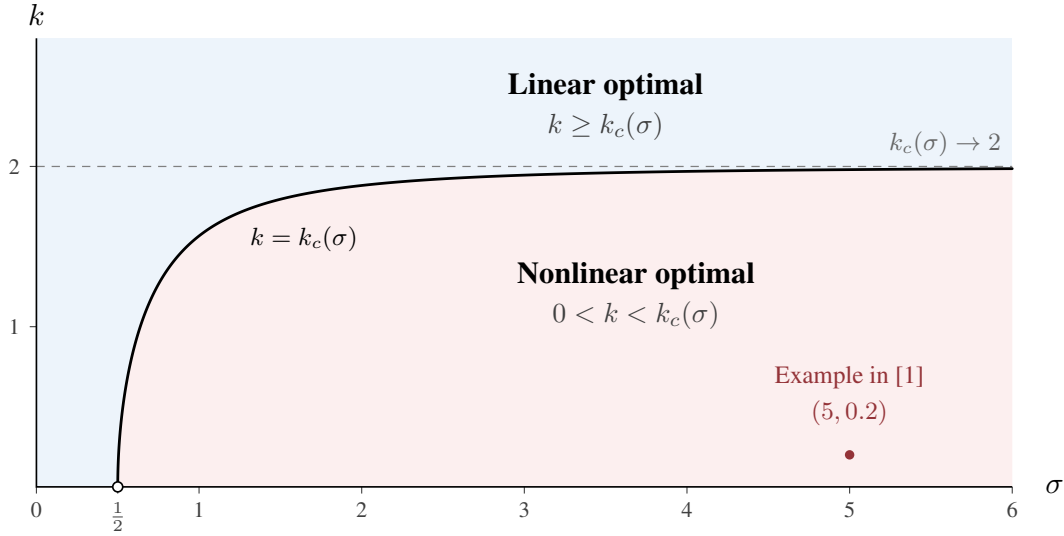}}
    \caption{\textbf{Phase transition} between linear and nonlinear optimal control for $k,\sigma>0$ (Theorem~\ref{thm_1}). The boundary is given by \cref{eq:intro-critical-k} and~\cref{eq:intro-critical-q}. Above and on the curve, the optimal controller pair is unique up to null sets and has the explicit linear form in~\cref{eq:intro-linear-pair}. Below it, every optimal first controller is nonlinear\protect\footnotemark[\value{footnote}]. The threshold is zero for $0<\sigma\le\frac12$ and approaches the dashed limit $k=2$ as $\sigma\to\infty$. The marked point is the classical example~\cite{counterexample} at $(\sigma,k)=(5,0.2)$. Appendix~\ref{app:numerical-validation} reports numerical tests of the boundary and the linear controller characterization.}
    \label{fig:critical-boundary}
\end{figure}

The critical curve $k=k_c(\sigma)$, defined by \cref{eq:intro-critical-k} and~\cref{eq:intro-critical-q} and shown in Figure~\ref{fig:critical-boundary}, separates linear optimality from nonlinear optimality. For $k\ge k_c(\sigma)$, the globally optimal controller pair is unique up to null sets and has the linear form in~\cref{eq:intro-linear-pair}. For $0<k<k_c(\sigma)$, every globally optimal first controller is nonlinear, even after modification on a set of Gaussian measure zero.

The nonlinear region raises a further structural question. A nonlinear policy may still have a limiting gain at large states and admit a simple description in the tails. However, current performance bounds and numerical constructions~\cite{GroverParkSahai2013, LeeLauHo2001, Mehmetoglu2014} do not determine whether every exact optimizer has such a description over the entire Gaussian support. We therefore study the normalized encoder gain $\frac{f(x)}x$ as $x \to \pm \infty$ and its implications for exact representation within a prescribed class of functions.

We use then tame geometry to make this representation question precise. A function is called \textit{tame} if it is definable, with real parameters, in an $o$-minimal expansion of the ordered real field~\cite{vandenDries1998}. This framework includes semialgebraic functions, and the $o$-minimal monotonicity theorem guarantees that every bounded definable function is eventually monotone on each tail. Consequently, any tame encoder with at most linear growth must have a convergent normalized gain on both tails.

Below the critical curve, every canonical optimal encoder has distinct lower and upper limiting gains on both tails and admits no tame representative. Here, the canonical encoder is a non-decreasing representative that is continuous from the right. The conclusion holds even for representations that agree with an optimal encoder only almost everywhere on a tail. Consequently, the phase transition between linear and nonlinear optimality also gives the exact boundary for tame representability.

\refstepcounter{footnote}
\footnotetext{More precisely, \emph{non-tame}; i.e., it is not definable in any $o$-minimal expansion of the ordered real field, as defined in Definition~\ref{def:tameness}.}

\subsection{Contributions}

Our main result gives the exact boundary between linear optimality and nontame optimal control over the positive parameter plane (Theorem~\ref{thm_1} and Corollary~\ref{cor_1}).

\begin{samepage}
The threshold is
\begin{subequations}
\label{eq:intro-boundary}
\begin{equation}
\label{eq:intro-critical-k}
k_c(\sigma)=
\begin{cases}
0, & 0<\sigma\le\frac12, \\
(2-q_\sigma)\sqrt{1-q_\sigma^2}, & \sigma>\frac12
\end{cases},
\end{equation}
where, for $\sigma>\frac12$, $q_\sigma\in(0,1)$ is the unique solution of the quartic equation
\begin{equation}
\label{eq:intro-critical-q}
\sigma^2q_\sigma(2-q_\sigma)(1+q_\sigma)^2=1.
\end{equation}
\end{subequations}
\end{samepage}
\cref{eq:intro-critical-q} is equivalent to $q_\sigma^4-3q_\sigma^2-2q_\sigma+\sigma^{-2}=0$, so a formula in radicals is available~\cite[\S1.11(iii)]{NISTDLMF}. The unique root characterization selects the required branch and permits direct evaluation by bisection. Thus, every positive control weight yields a unique linear optimum for $0<\sigma\le\frac12$. For $\sigma>\frac12$, the threshold increases toward $2$ as $\sigma\to\infty$.

For $k\ge k_c(\sigma)$, the globally optimal controller pair is unique up to null sets and is given by
\begin{equation}
\label{eq:intro-linear-pair}
f^*(x)=\alpha x,\qquad
 g^*(y)=\frac{\alpha^2\sigma^2}{1+\alpha^2\sigma^2}y,
\end{equation}
where $\alpha\in(0,1)$ is the unique solution of
\begin{equation}
\label{eq:intro-linear-gain}
k^2(1-\alpha)=\frac{\alpha}{(1+\alpha^2\sigma^2)^2}.
\end{equation}
Clearing the denominator in \cref{eq:intro-linear-gain} gives a quintic equation in $\alpha$. The Abel Ruffini theorem rules out a general formula in radicals for quintic equations~\cite{Alekseev2004}. Thus, the linear form in \cref{eq:intro-linear-pair} is explicit, while its gain is determined by a scalar equation whose unique root in $(0,1)$ can be computed numerically.

This characterization includes equality on the boundary. For $0<k<k_c(\sigma)$, every globally optimal first controller is nontame, and every canonical optimal encoder has distinct lower and upper limiting gains on each tail. In particular, a finite limiting gain on either tail is possible precisely in the linear region.

The analysis rests on a connection between a necessary condition derived from the tails and a sufficient condition for global optimality. We first establish an at most linear growth bound and show that any limiting gain must satisfy both a stationary equation and an inequality obtained by modifying the controller on a distant source tail (Lemmas~\ref{lem_2}, \ref{lem_3}, and~\ref{lem_4}). We then use the optimal transport formulation of Wu and Verd\'u~\cite{Wu2011} to compare arbitrary output laws with a Gaussian law. A Gaussian smoothing inequality bounds a combination of relative Fisher information and relative entropy by the squared Wasserstein distance (Lemmas~\ref{lem_5} and~\ref{lem_6}). Under the stationary equation, the resulting sufficient condition for global optimality is equivalent to the necessary tail condition. This agreement yields the exact boundary, while the entropy remainder establishes uniqueness on the boundary.

We also characterize when exact optimality is compatible with tame geometry (Corollary~\ref{cor_1}). Below the critical curve, no function definable in an $o$ minimal expansion can agree almost everywhere with an optimal encoder on either tail. Consequently, no family depending on finitely many real parameters and having an evaluation graph definable in such an expansion can contain an exact optimal encoder. The conclusion is invariant under modifications on sets of Gaussian measure zero and applies to the original first controller $\gamma_1^*(x)=f^*(x)-x$. This conclusion concerns exact representation and does not assert a positive performance gap for tame approximations.

We complement the analysis with numerical tests in Appendix~\ref{app:numerical-validation}. Direct searches of the original objective at $192$ parameter pairs find $39$ improvements over the best linear cost, all below the predicted boundary. Separately, numerical minimization over both linear gains at $5742$ parameter pairs agrees with \cref{eq:intro-linear-pair} and~\cref{eq:intro-linear-gain}, including equality on the boundary. Figures~\ref{fig:boundary-validation} and~\ref{fig:linear-validation} summarize these tests and the consistency of the boundary formulas.

\subsection{Related Work}

The relationship between information structure and linear optimality is fundamental to stochastic team theory. Ho and Chu~\cite{HoChu1972} established linear optimality for Gaussian quadratic teams with partially nested information. Witsenhausen's problem has a nonclassical information structure, since the first action affects the second observation without making the first controller's information available to the second. Bansal and Ba\c{s}ar~\cite{BansalBasar1987} studied a parameterized family of such teams and proved linear optimality when the cost contains no product term between the two decision variables. Their result concerns a structural condition on the cost. Our classification concerns variation of $k$ and $\sigma$ within the original Witsenhausen cost and information structure.

Information theoretic analyses emphasize the use of control actions for implicit communication. Mitter and Sahai~\cite{MitterSahai1999} showed that quantization based strategies can outperform the best linear strategy by an arbitrarily large factor over suitable parameter sequences. Grover, Park, and Sahai~\cite{GroverParkSahai2013} combined achievable strategies with lower bounds to approximate the optimal cost within a constant factor uniformly over the problem parameters. These results quantify the benefit of signaling and the performance of structured policies. The present analysis determines a property shared by every exact optimizer, including parameter values for which a small cost difference may give little information about the tails.

Numerical optimization provides detailed examples of nonlinear policy structure. Lee, Lau, and Ho~\cite{LeeLauHo2001} optimized piecewise linear policies and obtained an improved numerical solution. Mehmetoglu, Akyol, and Rose~\cite{Mehmetoglu2014} used deterministic annealing to obtain further improvements. Such computations are useful for identifying effective policy families and attainable costs. Our representability result addresses whether an exact optimizer can belong to a tame family on an unbounded tail, a conclusion that does not follow from numerical optimization over a prescribed representation.

The distributional approach is particularly relevant to our proofs. Wu and Verd\'u~\cite{Wu2011} expressed the objective as a quadratic Wasserstein transport cost plus a minimum mean square estimation error. They also established asymptotic affine optimality as the source variance tends to zero, while leaving exact affine optimality at fixed positive parameters unresolved. Zhu~\cite{Zhu2026} developed variational conditions and analyzed stability of a Gaussian branch under the second variation. That stability criterion does not establish global affine optimality. Our Gaussian comparison controls all output laws with finite second moment, and its agreement with the necessary tail condition gives the exact global boundary.

The geometric conclusion uses the finiteness and monotonicity properties of definable functions developed in tame geometry~\cite{vandenDries1998}. These properties imply convergence of a bounded definable gain on a tail. Their application here requires both a growth estimate for optimal encoders and a way to transfer tail convergence from an almost everywhere representative to the canonical monotone encoder. Combining these facts with the boundary theorem links definability to the parameters of the control problem.

\section{Necessary Condition}
\label{sec:main_results_1}

\textbf{Notational Convention.}
We denote $X_0 \sim \mathcal{P}$ and $X_1 \sim \mathcal{Q},$ where $\mathcal P = \mathcal N (0, \sigma ^2)$ and $\mathcal Q$ are probability distributions. We further use $ \varphi(x) = \frac{e ^{-\frac{1}{2}x^2}}{\sqrt{2 \pi}}$ and $\Phi$ for density function and distribution function of standard normal distribution, and $\varphi _\sigma (x) = \frac{ \varphi(\frac{x}{\sigma})}{\sigma} $ for the density function of $\mathcal{N}(0, \sigma^2).$ Moreover, the quadratic Wasserstein distance is defined as 
\begin{equation*}
    W_2 (\mathcal{P}, \mathcal{Q}) = \inf_{\mathbb P_{XY}} \{ \| X- Y\|_2: \mathbb P_X = \mathcal{P}, \mathbb P_Y = \mathcal{Q} \},
\end{equation*}
where $\| X- Y\|_2  \overset{\Delta}{=} \sqrt{\mathbb E[ (X - Y) ^2]}.$ Here $X$ and $Y$ are arbitrary random variables, and
\begin{equation*}
    \mathcal{R}_{\mathcal{Q}}(u) = \mathbb E_{N} [ (u - g_{\mathcal{Q}} (u + N) )^2 ].
\end{equation*}

Finally, we denote by $\mathcal W_2(\mathbb R)$ the space of probability distributions on $\mathbb R$ with finite second moments, equipped with the quadratic Wasserstein distance $W_2$.

We begin our proof by reducing the original problem to optimal transport shown in \cite{Wu2011}.

\begin{lemma}[optimal transport reduction in \cite{Wu2011}]
\label{lem_1}

The minimum in \cref{eq:cost} is equivalent to
\begin{equation}
    {J} ^*_{k, \sigma} = \min_{\mathcal Q} \{ k^2 W_2 ^2 (\mathcal{P}, \mathcal{Q}) + \text{mmse}(\mathcal{Q}, 1) \},\label{eq:optimal_transport}
\end{equation}
where $\text{mmse}(\mathcal Q, 1) = \mathbb E[(X_1-\mathbb E[X_1\mid Y])^2].$

\end{lemma}

We omit the proof of Lemma~\ref{lem_1} and refer the readers to \cite{Wu2011} for further details.

We next establish growth bounds for globally optimal controllers to constrain their asymptotic behavior and provide the foundation for subsequent posterior and tail analysis. Lemma~\ref{lem_2} aims to exploit the Gaussian observation structure to bound the MMSE decoder $g$ and its associated estimation cost, and then use global optimality to derive the point-wise best-response inequality for the encoder.

\begin{lemma}
\label{lem_2}

Suppose $\mathcal Q \in \mathcal{W}_2 (\mathbb R)$ and we denote $f ^*, g ^*$ as the optimal controller pair which minimizes \cref{eq:cost}. Then, there exists a constant $C < \infty$ such that
\begin{equation}
\label{eq:bound_f}
k^2 (f ^*(x) - x) ^2 \le \frac{k ^2}{1 + k^2} x ^2 + C|x| + C
\end{equation}
for any initial state realization $X_0 = x$.
\end{lemma}

\begin{proof}
For the fixed distribution $\mathcal Q$, define the partition function as
\begin{equation*}
\mathcal M(y) = \int_{u \in \mathbb{R}} e^{yu - \frac{u ^2}{2}} \mathcal Q (du),
\end{equation*}
and correspondingly, we define the log-partition function as
\begin{equation*}
\mathcal{L} (y) = \log \mathcal{M} (y).
\end{equation*}

For an arbitrary constant $0 < t < \infty$ and an arbitrary integer $l \in \{0, 1, 2\}$, we have
\begin{align}
\sup_{ |y| \le t, u \in \mathbb{R}} |u| ^l e^{y u - \frac{u ^2}{2}} & \le \sup_{ u \in \mathbb{R}} (1 + u ^2) e^{t |u| - \frac{u ^2}{2}} \nonumber \\
& \le \sup_{ u \in \mathbb{R}} (1 + u ^2) e^{t ^2 + \frac{u ^2}{4} - \frac{u ^2}{2}} \nonumber \\
& = \exp{(t ^2)} \sup_{ u \in \mathbb{R}} (1 + u ^2) e^{-\frac{u ^2}{4}} \nonumber \\
& \le \exp{(t ^2)} \left(1 + \frac{4}{e} \right) \nonumber \\
& < \infty,
\label{eq:lem2_eq1}
\end{align}
where the second inequality uses \emph{Young's inequality} and the third uses $\sup_{ u \in \mathbb{R}} u ^2 \exp{({-\frac{u ^2}{4}})} \le \frac{4}{e}$. Since $\mathcal Q$ is a probability measure for $X_1$, the uniform bound above provides a \(\mathcal Q\)-integrable dominating function on every compact interval of \(y\). Furthermore, by the \emph{mean value theorem}, the difference quotients corresponding to the first two derivatives are bounded by the same type of integrable constant. Therefore, the \emph{dominated convergence theorem} justifies that $0 < \mathcal{M}(y) < \infty$ is twice continuously differentiable, and so does $\mathcal{L} (y).$

Conditional on $X_1 = u,$ the observation \(Y\) has a density of
\begin{equation*}
    \varphi (y - u) = \frac{e ^{-\frac{1}{2} y ^2 + yu - \frac{1}{2}u ^2}}{\sqrt{2\pi}},
\end{equation*}
and therefore the marginal density of \(Y\) is
\begin{align}
p_{\mathcal Q} (y) & = \int_{u \in \mathbb R} \varphi (y-u) \mathcal Q(du) \nonumber \\
& = \frac {1}{ \sqrt{2 \pi}} \int_{ u \in \mathbb R } \exp \left( - \frac{(y - u) ^2}{2} \right) \mathcal Q(du) \nonumber \\
& = \frac{1}{ \sqrt{ 2 \pi}} e^{-\frac{y ^2}{2}} \int_{ u \in \mathbb R} e^{yu -\frac{u ^2}{2}} \mathcal Q(du) \nonumber \\
& = \frac{1}{\sqrt {2 \pi}} e^{-\frac{y ^2}{2}} \mathcal M(y). \nonumber
\end{align}

Since \(p_{\mathcal Q}(y) > 0\) for every \(y \in \mathbb R,\) \emph{Bayes rule} yields the following distribution of \(X_1\) given \(Y = y\)
\begin{align}
\mathbb P_{\mathcal Q} (X_1 \in du \mid Y = y) & = \frac{\varphi (y - u) \mathcal Q (du)} {p_{\mathcal Q}(y)} \nonumber \\
& = \frac{ \frac{1}{\sqrt{2 \pi}} e^{-\frac{y ^2}{2} + yu -\frac{u^2}{2}} } { \frac{1}{\sqrt{2 \pi}} e^{-\frac{y ^2}{2}} \mathcal M (y) } \mathcal Q(du) \nonumber \\
& = \frac{e^{yu - \frac{u ^2}{2}}}{\mathcal M (y)} \mathcal Q (du),
\label{eq:lem2_eq2}
\end{align}
where $\mathcal Q$ does not need to admit a Lebesgue density.

Therefore, applying \cref{eq:lem2_eq2}, the optimal \emph{minimum mean square error (MMSE)} decoder satisfies
\begin{align}
g_{\mathcal Q} (y) & = \mathbb E_{\mathcal Q} [X_1 \mid Y=y] \nonumber \\
& = \int_{ u \in \mathbb R} u \mathbb P _{\mathcal Q} (X_1 \in du \mid Y = y) \nonumber \\
& = \frac{\int_{ u \in \mathbb R} u e^{yu-\frac{u^2}{2}} \mathcal Q (du)} {\mathcal M (y)} \nonumber \\
& = (\log \mathcal{M} (y))' \nonumber \\
& = \mathcal{L}' (y).
\end{align}

Similarly, differentiating \(\mathcal L' (y) = \frac{\mathcal M'(y)} {\mathcal M(y)} \) gives
\begin{align}
\mathcal L''(y) & = \frac{\mathcal M''(y) \mathcal M(y) - \mathcal M'(y) ^2}{ \mathcal M(y) ^2} \nonumber \\
& = \frac{\mathcal M''(y)}{ \mathcal M (y)} - \left( \frac{\mathcal M'(y)}{ \mathcal M(y)} \right) ^2 \nonumber \\
& = \int_{ u \in \mathbb R} u^2 \mathbb P _ {\mathcal Q} (X_1 \in du \mid Y = y) - \left( \int_{ u \in \mathbb R}u \mathbb P _ {\mathcal Q}(X_1 \in du \mid Y = y) \right) ^2 \nonumber \\
& = \mathbb E_{\mathcal Q} [X_1^2 \mid Y = y] - \mathbb E_{\mathcal Q}[X_1 \mid Y = y] ^2 \nonumber \\
& = \mathrm{Var}_{\mathcal Q} (X_1 \mid Y = y) \nonumber \\
& \ge 0.
\label{eq:lem2_eq3}
\end{align}

Hence, \( \mathcal L\) is convex on \(\mathbb R\).

As $n$ increases, the set $[-n, n]$ increases to $\mathbb{R}$, hence $\lim_{n \to \infty}\mathcal Q ([-n, n]) = 1$.
We then choose $0 < A < \infty$ with $m \coloneqq \mathcal Q ([-A, A]).$ We now lower and upper bound $\mathcal M(y),$ respectively
\begin{gather*}
\mathcal M (y) = \int_{u \in \mathbb R} e^{\frac{y ^2}{2} - \frac{(u-y)^2}{2}} \mathcal Q (du) \le e ^{\frac{y^2}{2}},
\\
\mathcal M(y) \ge \int_{u \in [-A, A]} e^ {yu - \frac{u^2}{2}} \mathcal Q (du) \ge m e ^{-A |y|- \frac{A^2}{2}}.
\end{gather*}

Taking logarithms to the above equations gives
\begin{equation}
\label{eq:lem2_eq4}
\log m - A|y| - \frac{A ^2}{2} \le \mathcal L (y) \le \frac{y ^2}{2}.
\end{equation}

For any $y \ge 1,$ convexity of $\mathcal L$ yields the two secant comparisons
\begin{equation*}
\frac{ \mathcal L (y) - \mathcal L(0)} {y} \le \mathcal L'(y) \le \frac{ \mathcal L (2y) - \mathcal L (y)} {y}.
\end{equation*}

Combining \cref{eq:lem2_eq4} and \cref{eq:lem2_eq3} gives
\begin{equation*}
-A+ \frac{\log m - \frac{A^2}{2} - \mathcal L (0)} {y} \le g_{\mathcal Q} (y) \le 2y + A + \frac{\frac{A ^2}{2} - \log m}{y}.
\end{equation*}

Hence there exists a finite $C$ such that 
\begin{equation*}
    -C \le g_{\mathcal Q} (y) \le 2y + C
\end{equation*}
for all $y \in [1, \infty).$ We also apply the same argument to the reflected law to obtain
\begin{equation*}
2y - C \le g_{\mathcal Q} (y) \le C
\end{equation*}
for all $y \in ( -\infty, -1]$.

Since $g_{\mathcal{Q}} (y)$ is twice continuously differentiable in $[-1, 1]$, we obtain
\begin{equation}
|g_{\mathcal Q} (y)| \le 2|y| + C.
\label{eq:lem2_eq5}
\end{equation}

Consequently, applying triangle inequality to \cref{eq:lem2_eq5} gives
\begin{align*}
|u - g_{\mathcal Q} (u + N) | & \le | (u + N) - g_{\mathcal Q} (u + N)| + |N| \nonumber \\
& \le |u| + 2|N| + C.
\end{align*}

Therefore, define $\mathcal{R}_{\mathcal{Q}} (u) = \mathbb E_{N} [ (u - g_{\mathcal{Q}} (u + N)) ^2]$ and the squared expectation is at most
\begin{align}
\mathcal{R}_{\mathcal{Q}} (u) & = \mathbb E_{N} [ (u - g_{\mathcal{Q}} (u + N)) ^2] \nonumber \\
& \le \mathbb E_{N} [ ( |u| + 2|N| + C) ^2] \nonumber \\
& = \mathbb E_{N} [u ^2 + 2|u| (2 |N| + C) + (2|N| + C) ^2] \nonumber \\
& = u ^2 + 2 |u| \mathbb{E} [2 |N| + C] + \mathbb{E} [(2 |N| + C) ^2] \nonumber \\
& = u ^2 + (4 \mathbb E [|N|] + 2C ) |u| + (4 \mathbb E [N ^2] + 4C \mathbb E [|N|] + C ^2) \nonumber \\
& = u^2+C_1|u|+C_2,
\label{eq:lem2_eq6}
\end{align}
where $C_1 = 4 \sqrt{\frac{2}{ \pi}} + 2 C$ and $C_2 = 4 + 4C \sqrt{\frac{2}{ \pi}} + C ^2$ since $N \sim \mathcal{N} (0, 1).$

In \cref{eq:lem2_eq6}, dividing both sides by $u ^2$ and taking $\limsup$ as \(|u| \to \infty \), we obtain
\begin{align}
\limsup_{|u| \to \infty} \frac{\mathcal{R}_{\mathcal Q} (u)}{ u ^2} & \le \lim_{|u| \to \infty} (1 + \frac{C_1}{|u|} + \frac{C_2}{u ^2}) \nonumber \\
& = 1.
\label{eq:lem2_eq7}
\end{align}

Moreover, we can easily justify that $\mathcal{R}_{\mathcal{Q}} (u)$ is continuous by \emph{dominated convergence theorem.}

We fix an optimal pair of controllers \((f ^*, g ^*)\), where \( g ^*\) is the smooth \emph{MMSE} decoder corresponding to the optimal output distribution $\mathcal{Q}^*.$ For every \(x, u \in \mathbb R \), we define the conditional cost
\begin{equation*}
\mathcal C (x, u) \coloneqq k^2 (u -x) ^2 + \mathcal R_{\mathcal Q^*} (u).
\end{equation*}

By the definition of optimal cost,
\begin{align}
J_{k, \sigma} ^* & = J_{k, \sigma} (f ^*, g ^*) \nonumber \\
& = k ^2 \mathbb E[ (f ^* (X_0) - X_0) ^2] \label{eq:lem2_eq8} \\
& + \mathbb E[ (f ^*(X_0) - g^*(f^*(X_0) + N)) ^2]. \label{eq:lem2_eq9}
\end{align}

Since \( X_0 \) and \( N \) are independent, applying \emph{Tonelli's theorem} to \cref{eq:lem2_eq9} yields
\begin{align}
& \mathbb E[(f ^*(X_0) - g^*( f^*(X_0) + N)) ^2] \nonumber \\
& = \int_{z \in \mathbb R} \int_{x \in \mathbb R} (f ^*(x) - g^*( f^*(x) + z)) ^2 \varphi (z) \varphi_\sigma (x) dz dx \nonumber \\
& = \int_{x \in \mathbb R} \mathcal R_{ \mathcal Q^*} (f ^*(x)) \varphi_ \sigma(x) dx.
\label{eq:lem2_eq10}
\end{align}

Consequently,
\begin{align}
J_{k, \sigma} ^* & =\int_{x \in \mathbb R} \left( k ^2( f^*(x) - x) ^2 + \mathcal R_{\mathcal Q^*}(f ^*(x)) \right) \varphi_\sigma(x) dx \nonumber \\
& = \int_{x \in \mathbb R} \mathcal C(x, f^*(x) ) \varphi_\sigma(x) dx.\label{eq:lem2_eq11}
\end{align}

The same identity holds for any Borel measurable competing encoder \( \tilde f\)
\begin{equation*}
J_{k, \sigma} (\tilde f, g^*) = \int_{x \in \mathbb R} \mathcal C(x, \tilde f(x)) \varphi_\sigma (x)dx.
\end{equation*}

For each rational number \(r\) and positive integer \(t\), we construct a helper set $\mathcal E$ as
\begin{equation*}
\mathcal E_{r, t} = \left \{ x \in \mathbb R: \mathcal C (x, r) \le \mathcal C (x, f^*(x)) - \frac{1}{t} \right \}.
\end{equation*}

Since \(f ^*\) is Borel measurable and \( \mathcal R_{\mathcal Q^*}\) is continuous, \(\mathcal E_{r, t}\) is a Borel measurable.

Suppose, toward a contradiction, that
\begin{equation*}
    \mathbb P (X_0 \in \mathcal{E}_{r, t}) > 0.
\end{equation*}

We define a competing encoder $\tilde f$ as
\begin{align*}
\tilde f(x) = \begin{cases} r, & x \in \mathcal E_{r, t}, \\
f ^*(x), & \mathrm{otherwise}
\end{cases}.
\end{align*}

As \(\mathcal E_{r, t}\) is Borel measurable and \(f ^*\) is Borel measurable, \(\tilde f\) is also Borel measurable. Moreover, by the definition of \(\mathcal E_{r, t}\), we have
\begin{equation*}
\mathcal C (x, \tilde f(x) ) \le \mathcal C (x, f^*(x) ) \qquad \text{for every }x \in \mathbb R.
\end{equation*}

Since \(\tilde f(x) = f ^*(x)\) for every \(x \notin \mathcal E_{r, t}\), the integrand vanishes outside \(\mathcal E_{r, t}\). Therefore,
\begin{align}
& J_{k, \sigma} (\tilde f, g ^*) - J_{k, \sigma} ^* \nonumber \\
& = \int_{x \in \mathcal E_{r, t}} \big( \mathcal C(x, r) - \mathcal C(x, f ^*(x)) \big) \varphi_\sigma(x) dx \nonumber \\
& \le -\frac {1}{t} \int_{x \in \mathcal E_{r, t}} \varphi_ \sigma(x) dx \nonumber \\
& = -\frac{1}{t} \mathbb P(X_0 \in \mathcal E_{r, t}) \nonumber \\
& < 0, \nonumber
\end{align}
which contradicts to the optimality of $f ^*, g ^*.$ Hence
\begin{equation}
\mathbb P (X_0 \in \mathcal{E}_{r, t}) = 0
\label{eq:lem2_eq12}
\end{equation}
for all rational number $r$ and all positive integer $t.$

We then define the exceptional null set as the union of $\mathcal{E}_{r, t}$. Formally, 
\begin{equation*}
\mathcal E = \bigcup_{r \in \mathbb Q} \bigcup_{t=1}^{ \infty} \mathcal E_{r, t}.
\end{equation*}

Since \(\mathbb Q\) and \(\mathbb N\) are countable, \(\mathcal E\) is a countable union of null sets. By countable sub-additivity,
\begin{equation*}
\mathbb P (X_0 \in \mathcal E) = 0.
\end{equation*}

Fix an arbitrary \(x \in \mathbb R \setminus \mathcal E\). For every rational \(r\) and \(t \ge 1\), we have \(x \notin \mathcal E_{r, t}\), which implies
\begin{equation*}
\mathcal C(x, f ^*(x)) < \mathcal C(x, r) + \frac{1}{t}.
\end{equation*}

Letting \(t \to \infty \), we obtain
\begin{equation}
\mathcal C(x, f ^*(x)) \le \mathcal C(x, r) \qquad \forall r \in \mathbb Q.
\label{eq:lem2_eq13}
\end{equation}

Thus this best-response inequality holds simultaneously for every rational competing action on the same full-measure set \(\mathbb R \setminus \mathcal E\).

Fix any \(x \in \mathbb R \setminus \mathcal E\) and \(u \in \mathbb R\). There exists a sequence \((r_l)_ {l \ge 1} \subseteq \mathbb Q\) such that $r_l \to u.$ By \cref{eq:lem2_eq13},
\begin{equation*}
\mathcal C(x, f ^*(x)) \le \mathcal C(x,r_l) \qquad \forall l \ge 1.
\end{equation*}

Since $\mathcal{C} (x, u)$ is continuous with respect to $u$, we have
\begin{equation*}
\lim_{l \to \infty} \mathcal C(x, r_l) = \mathcal C(x, u).
\end{equation*}

Taking \(l \to \infty\) yields
\begin{equation*}
\mathcal C(x, f ^*(x)) \le \mathcal C(x, u).
\end{equation*}

Since \(x \in \mathbb R \setminus \mathcal E\) and \(u \in \mathbb R\) are arbitrary, we conclude that
\begin{equation}
k ^2 (f ^*(x) - x) ^2 + \mathcal R_{\mathcal Q ^*} (f ^*(x)) \le k ^2(u - x) ^2 + \mathcal R_{\mathcal Q^*}(u)
\label{eq:lem2_eq14}
\end{equation}
for all $u \in \mathbb R$ and for \emph{almost every} \(x \in \mathbb R\).

We then generalize this property for every $x$ by choosing $u = a x$, where $a = \frac{k ^2}{1+k^2}$. Combining \cref{eq:lem2_eq14} and \cref{eq:lem2_eq6},
\begin{align}
k ^2( f ^*(x) - x) ^2 & \le k ^2( a - 1 ) ^ 2x ^2 + \mathcal R_{\mathcal Q^*} (ax) \nonumber \\
& \le k ^2 (a-1) ^2 x^2 + a ^2 x ^2 + C_1 a|x| + C_2 \nonumber \\
& = \frac{k ^2}{1 + k^2}x ^2 + C_1 a |x| + C_2 \nonumber \\
& \le \frac{k ^2}{1 + k ^2}x ^2 + C|x| + C
\label{eq:lem2_eq15}
\end{align}
for a sufficiently large $C.$

The inequality holds on the common full-measure set \(\mathbb R \setminus \mathcal E\). Since the Gaussian density \(\varphi_\sigma\) is strictly positive on \(\mathbb R\), this set is dense in \(\mathbb R\). For any \(x \in \mathbb R\), we choose a sequence \((x_n)_{n \ge 1} \subseteq \mathbb R \setminus \mathcal E\) such that \(x_n \downarrow x\). By the right-continuity of the canonical encoder \(f^*\),
\begin{equation*}
    \lim_{n \to \infty}f ^*(x_n) = f^*(x).
\end{equation*}

Applying \cref{eq:lem2_eq15} at each \(x_n\) and letting \(n \to \infty\), we obtain
\begin{equation}
k ^2 (f ^*(x) - x) ^2 \le \frac{k ^2}{1 + k^2} x^2 + C|x| + C
\label{eq:lem2_eq16}
\end{equation}
for \emph{every} \(x \in \mathbb R\).

Finally, dividing \cref{eq:lem2_eq16} by \(k ^2 x^2\) for any \(x \neq 0\), we obtain
\begin{equation*}
\left ( \frac{f ^*(x)} x - 1 \right) ^2 \le \frac{1}{1 + k^2} + \frac{C}{k ^2|x|} + \frac {C}{k ^2 x ^2}. 
\end{equation*}

Letting \(x \to +\infty\), we obtain
\begin{align*}
1 - \frac{1}{\sqrt{1 + k ^2}} & \le \liminf_{x \to + \infty} \frac{f ^*(x)} x \\
& \le \limsup_{x \to +\infty} \frac{f ^*(x)} x \\
& \le 1 + \frac{1}{\sqrt{1+k^2}}.
\end{align*}

The same argument applies as \(x \to -\infty \), since the preceding inequality holds for every \(x \neq 0\).

This completes the proof of \Cref{lem_2}.

\end{proof}

\paragraph{Consequences of Lemma~\ref{lem_2}.} A natural consequence of \cref{eq:lem2_eq16} can also be derived by picking a sufficiently large constant $d$ such that the following inequalities hold simultaneously
\begin{equation*}
2 \frac{1}{\sqrt{1 + k^2}} d \ge \frac{C}{k ^2}, \qquad d^2 \ge \frac{C}{k ^2}.
\end{equation*}

We can then obtain
\begin{align*}
(f ^*(x) - x) ^2 & \le \frac{1}{1 + k ^2}x^2 + \frac{C}{k ^2}|x| + \frac{C}{k ^2} \\
& \le \frac{1}{1 + k ^2} x^2 + 2d \frac{1}{\sqrt{1 + k^2}} |x| + d ^2 \\
& = \left(\frac{1}{\sqrt{1 + k^2}} |x| + d \right)^2.
\end{align*}

Taking square root and using triangle inequality give
\begin{align}
|f ^*(x)| & \le |f ^*(x) - x| + |x| \nonumber \\
& \le (1 + \frac{1}{\sqrt{1 + k^2}}) |x| + d \nonumber \\
& \le 2|x| + d.
\label{eq:lem2_eq17}
\end{align}

The preceding bounds ensure that $\frac{f^*(x)}x$ is bounded above and away from zero for sufficiently large $x$. Lemma~\ref{lem_3} shows that, whenever this ratio converges, its limit must lie in $(0, 1)$ and satisfy a scalar stationary  condition, derived from the decoder and estimation-cost asymptotics.

\begin{lemma}
\label{lem_3}
Fix a canonical optimal controller pair $(f^*, g^*)$. Suppose the limit $\lim_{x \to + \infty} \frac{f^*(x)}{x} = \alpha$ exists and is finite. Then $\alpha$ satisfies the following stationary condition
\begin{equation}
\label{eq:alpha_condition}
k ^2 (1 - \alpha) = \frac{ \alpha}{ (1 + \alpha ^2 \sigma ^2) ^2}.
\end{equation}
Moreover, we can show that $\alpha \in (0, 1).$
\end{lemma}

\begin{proof}
Since $X_0 \sim \mathcal N(0, \sigma^2)$ and $Y = f^*(X_0) + N$ with $N \sim \mathcal N(0, 1)$ independent of $X_0$, Bayes' rule gives, for any $y>0$, the conditional density
\begin{align}
p_{X_0 \mid Y}(x \mid y) & = \frac{\varphi _\sigma(x) \varphi (y - f ^*(x))} {\int_{s \in \mathbb R} \varphi_\sigma(s) \varphi(y - f ^*(s)) ds} \nonumber \\
& = \frac{ \exp \left(- \frac{x ^2}{2 \sigma ^2} - \frac{(y - f ^*(x)) ^2}{2} \right)} {\int_{s \in \mathbb R} \exp \left( - \frac{s ^2}{2 \sigma ^2} - \frac{(y - f ^*(s)) ^2}{2} \right) ds}.
\label{eq:lem3_eq1}
\end{align}

We now use change of variables $x = y t$ and denote the random variable $T_y = \frac{X_0}{y}$. For $y > 0$, the Jacobian factor $y$ cancels between the numerator and denominator. Consequently,
\begin{align}
p_{T_y \mid Y}(t \mid y) & = y p_{X_0 \mid Y} (yt \mid y) \nonumber \\
& = \frac{ y \exp \left( - \frac{y ^2 t ^2}{2 \sigma ^2} -\frac{(y - f ^*(yt)) ^2}{2} \right) }{ \displaystyle \int_{s \in \mathbb R} \exp \left( - \frac{s ^2}{2 \sigma^2} -\frac{(y - f ^*(s)) ^2}{2} \right) ds } \nonumber \\
& = \frac{ y \exp \left( -y ^2 \left( \frac{t ^2}{2 \sigma ^2} + \frac{1}{2} \left( 1 - \frac{f ^*(yt)} {y} \right) ^2
\right) \right) }{ \displaystyle y \int_{z \in \mathbb R} \exp \left( -y ^2 \left( \frac{z ^2}{ 2 \sigma ^2} + \frac{1}{2} \left(1 - \frac{f ^*(yz)} y \right) ^2 \right) \right) dz } \nonumber \\
& = \frac{e ^{-y ^2I_y(t)}} {\displaystyle\int_{z \in \mathbb R} e^{-y ^2I_y(z)} dz},
\label{eq:lem3_eq2}
\end{align}
where, for this proof only,
\begin{equation*}
I_y(t) = \frac{t ^2}{2 \sigma ^2} + \frac{1}{2} \left(1 - \frac{f ^*(yt)}{y} \right) ^2.
\end{equation*}

We then show uniform convergence of $\frac{f ^*(yt)}{y}$ on an arbitrary compact interval \( t \in [0, m]\), where \(m > 0\) is fixed. Suppose \(\varepsilon > 0\), then there exists a constant \( x_c > 0\) such that
\begin{equation*}
|f ^*(x) - \alpha x| \le \varepsilon x \qquad \text{for all } x \ge x_c.
\end{equation*}

The canonical encoder is locally bounded by Lemma~\ref{lem_2}. Hence
\begin{equation*}
\mathcal U_{x_c} \coloneqq \sup_{0 \le x \le x_c} |f ^*(x) - \alpha x| < \infty.
\end{equation*}

For any $yt \ge x_c$, we have
\begin{align*}
\left| \frac{f ^* (yt)}{y} - \alpha t \right| & = \frac{|f ^* (yt) - \alpha yt|}{y} \\
& \le \frac{ \varepsilon yt}{y} \\
& = \varepsilon t \\
& \le \varepsilon m,
\end{align*}
while \(0 \le yt < x_c\), by the definition of \(\mathcal U_{x_c}\), we have
\begin{align*}
\left| \frac{f ^* (yt)}{y} - \alpha t \right| & = \frac{|f ^* (yt) - \alpha yt|}{y} \\
& \le \frac{\mathcal U_{x_c}}{y}.
\end{align*}

Combining the two cases gives
\begin{align}
\sup_ {0 \le t \le m} \left| \frac{f ^*(yt)}{y} - \alpha t \right| & \le \max \left \{\varepsilon m, \frac{ \mathcal U_{x_c}} {y} \right\} \nonumber \\
& \le \varepsilon m + \frac { \mathcal U_{x_c}} {y}.
\label{eq:lem3_eq3}
\end{align}

Since \(\mathcal U_{x_c} < \infty\), taking \(y \to +\infty\) yields
\begin{equation*}
\limsup_{y \to +\infty} \sup_{0 \le t \le m} \left| \frac{f ^*(yt)}{y} - \alpha t \right| \le \varepsilon m.
\end{equation*}

As \(\varepsilon > 0\) is arbitrary, letting \(\varepsilon \downarrow 0\) gives
\begin{equation}
\label{eq:lem3_eq4}
\lim_{y \to +\infty} \sup_{0 \le t \le m} \left| \frac{f ^*(yt)}{y} - \alpha t \right| = 0.
\end{equation}

Indeed, the difference between the two rate functions is
\begin{align*}
|I_ y(t) - I_ \alpha(t)| & = \left| \frac{t ^2} {2 \sigma^2} + \frac{1}{2} \left(1 - \frac{f ^*(yt)}{y} \right) ^2 - \frac{t ^2}{2 \sigma ^2} - \frac{1}{2}(1 - \alpha t) ^2 \right| \\
& = \frac{1}{2} \left| \left(1 - \frac{f^*(yt)}{y} \right) ^2 - ( 1 - \alpha t) ^2 \right| \\
& = \frac{1}{2} \left| \left( 1 - \frac{f ^*(yt)}{y} - 1 + \alpha t \right) \left(1 - \frac{f ^*(yt)}{y} + 1 - \alpha t \right) \right| \\
& = \frac{1}{2} \left| \alpha t - \frac{f^*(yt)}{y} \right| \left| 2 - \frac{f^*(yt)}{y} - \alpha t \right| \\
& = \frac{1}{2} \left| \frac{f ^*(yt)}{y} - \alpha t \right| \left| 2 - \frac{f^*(yt)}{y} -\alpha t \right|.
\end{align*}

The first factor converges uniformly to zero by \cref{eq:lem3_eq4}, and the second factor is uniformly bounded on the compact interval $[0, m]$. Consequently, $I_y(t)$ converges uniformly on the interval $[0, m]$ to
\begin{align}
I_ \alpha(t) & = \frac{t^2}{2 \sigma^2} + \frac{(1 - \alpha t) ^2}{2} \nonumber \\
& = \frac{1 + \alpha ^2 \sigma^2}{2 \sigma^2} t^2 - \alpha t + \frac{1}{2} \nonumber \\
& = \frac{1 + \alpha ^2 \sigma ^2}{2 \sigma^2} \left(t - \frac{ \alpha \sigma^2}{1 + \alpha ^2 \sigma ^2} \right)^2 + \frac{1}{2 (1 + \alpha^2 \sigma^2)}.
\label{eq:lem3_eq5}
\end{align}

Since $\alpha > 0$, the quadratic function has the unique minimizer and minimum
\begin{equation}
t_* = \frac{ \alpha \sigma^2}{1 + \alpha ^2 \sigma ^2} > 0, \qquad I_* = \frac{1}{2 (1 + \alpha ^2 \sigma ^2)} < \frac{1}{2}.
\label{eq:lem3_eq6}
\end{equation}

Next, we show that conditional on a large observation $Y = y$, the normalized source $T_y$ concentrates at $t_*$. For $0 < \varepsilon < \frac{t_*}{2}$, this suffices for arbitrary positive $\varepsilon$, since the event of being farther than $\varepsilon$ from $t_*$ decreases as $\varepsilon$ increases. We then choose $\delta > 0$ small enough such that
\begin{equation}
5 \delta < \frac{1}{2} - I_*, \qquad 5 \delta < \frac{1 + \alpha ^2 \sigma ^2}{2 \sigma^2} \varepsilon ^2.
\label{eq:lem3_eq7}
\end{equation}

We then choose $A > t_*+\varepsilon+1$ sufficiently large such that
\begin{equation*}
\frac{A ^2}{4\sigma^2}>I_*+3\delta.
\end{equation*}

Since $I_\alpha(t)$ is continuous and attains its minimum $I_*$ at $t_*$, there exists a closed interval $\mathcal{I} \subseteq (0, A)$ of positive length containing $t_*$ such that
\begin{equation*}
I_\alpha(t) \le I_* + \frac{\delta}{2}
\qquad \forall t \in \mathcal I.
\end{equation*}

Since the uniform convergence established earlier holds on every fixed compact interval, it also holds on $[0, A]$. Therefore, for all sufficiently large $y$, we have
\begin{equation*}
|I_y(t)-I_\alpha(t)| \le \frac{\delta}{2}
\qquad \forall t \in [0, A].
\end{equation*}

Consequently, for every $t \in \mathcal I$,
\begin{align*}
I_ y(t) & \le I_ \alpha(t) + \frac{\delta}{2} \\
& \le I_* + \frac{ \delta}{2} + \frac{ \delta}{2} \\
& = I_* + \delta.
\end{align*}

Hence, the normalizing denominator satisfies
\begin{align}
\int_{t \in \mathbb R} e^ { -y ^2 I_y(t) } dt & \ge \int_{t \in \mathcal I} e^ { -y ^2 I_y(t) } dt \nonumber \\
& \ge \int_{t \in \mathcal I} e^ {-( I_* + \delta ) y^2} dt \nonumber \\
& = |\mathcal I| e^ {- (I_* + \delta) y^2 }.
\label{eq:lem3_eq8}
\end{align}

We now bound the posterior mass away from $t_*$ in three regions. On the interval $[0, A] \setminus (t_*-\varepsilon, t_*+\varepsilon)$, \cref{eq:lem3_eq5} and \cref{eq:lem3_eq7} gives
\begin{align}
I_\alpha(t) - I_* & = \frac {1 + \alpha^2 \sigma^2}{2 \sigma^2} (t - t_*) ^2 \nonumber \\
& \ge \frac{1 + \alpha^2 \sigma^2}{2 \sigma^2} \varepsilon^2 \nonumber \\
& > 5 \delta.
\label{eq:lem3_eq9}
\end{align}

By uniform convergence on $[0, A]$, we eventually have $|I_y(t) -I_\alpha(t)| \le \delta$ throughout this interval. Thus
\begin{align*}
I_y(t) & \ge I_\alpha(t) - \delta \\
& > I_* + 5\delta - \delta \\
& = I_* + 4\delta.
\end{align*}

On $[-A, 0]$, the monotonicity of the canonical encoder gives $f ^*(yt) \le f^*(0)$. For all sufficiently large $y$ such that $1 - \frac{f ^*(0)}{y} > 0$, it follows that
\begin{align}
I_y(t) & = \frac{t^2}{2 \sigma^2} + \frac{1}{2} \left(1 - \frac{f ^*(yt)}{y} \right) ^2 \nonumber \\
& \ge \frac{t^2}{2 \sigma^2} + \frac{1}{2} \left(1 - \frac{f ^*(0)}{y} \right) ^2 \nonumber \\
& \ge \frac{1}{2} \left(1 - \frac{f ^*(0)}{y} \right)^2.
\label{eq:lem3_eq10}
\end{align}

Since
\begin{align*}
\lim_{y \to +\infty} \frac{1}{2} \left(1 - \frac{f ^*(0)}{y} \right) ^2 & = \frac{1}{2} \\
& > I_* + 5 \delta,
\end{align*}
where the inequality follows from
\cref{eq:lem3_eq7}, we obtain
\begin{equation*}
I_ y(t) \ge I_* + 4 \delta \qquad \forall t \in [-A, 0]
\end{equation*}
for all sufficiently large $y$.

Combining these two bounded regions and using $1 + t^2 \le 1 + A ^2$, we obtain
\begin{equation}
\int_{ |t| \le A} (1 + t^2) \mathbbm{1} _{\{ | t - t_*| > \varepsilon \}} e ^{-y^2 I_y(t) }dt \le 2A (1 + A^2) e^ {-(I_* + 4 \delta)y^2 }.
\label{eq:lem3_eq11}
\end{equation}

It remains to control the unbounded region $|t| > A$. By the definition of $I_y(t)$, we have
\begin{align*}
I_y(t) & = \frac{t ^2}{2 \sigma^2} + \frac{1}{2} \left( 1 - \frac{f ^*(yt)}{y} \right) ^2 \\
& \ge \frac{t^2}{2\sigma^2}.
\end{align*}

Hence, splitting the Gaussian exponent into two equal parts, we obtain
\begin{align}
& \int_{|t| > A}(1 + t^2) \exp \big(-y^2 I_y(t) \big) dt \nonumber \\
& \le \int_{|t| > A}(1 + t^2) \exp \left(-\frac{y^2 t^2}{2 \sigma^2} \right) dt \nonumber \\
& = \int_{|t| > A}(1 + t^2) \exp \left(-\frac{y^2 t^2}{ 4 \sigma^2} \right) \exp \left(-\frac{y^2 t^2}{4 \sigma^2} \right) dt \nonumber \\
& \le \exp \left(- \frac{A^2 y^2}{4 \sigma^2}\right) \int_{t\in\mathbb R}(1 + t^2) \exp \left(- \frac{y^2 t^2}{4 \sigma^2} \right) dt \nonumber \\
& = \exp \left( -\frac{A^2 y^2}{4 \sigma^2} \right) \left( \frac{2 \sigma \sqrt\pi}{y} + \frac{4 \sigma^3 \sqrt \pi}{y^3} \right) \nonumber \\
& \le \frac{C_ \sigma}{y} \exp \left(-\frac{A^2 y^2}{4 \sigma^2} \right),
\label{eq:lem3_eq12}
\end{align}
where $C_\sigma = 2\sigma \sqrt\pi + 4\sigma^3 \sqrt \pi$ and $y \ge 1$.

Using the conditional density in \cref{eq:lem3_eq2} and the lower bound in \cref{eq:lem3_eq8},
\begin{align}
& \mathbb E \left[ (1 + T_y^2) \mathbbm 1 _{\{|T_y - t_*| > \varepsilon\}} \mid Y = y \right] \nonumber \\
& = \frac{ \displaystyle \int_{|t - t^*| > \varepsilon} (1 + t^2) e^ {-y ^2 I_y(t) } dt }{ \displaystyle\int_{t \in \mathbb R} e^ { -y ^2 I_y(t) } dt } \nonumber \\
& \le \frac{2A (1 + A ^2)}{ |\mathcal I|} \exp (-3 \delta y ^2 ) + \frac{C_\sigma}{|\mathcal I| y} \exp \left[ - \left( \frac{A ^2}{4 \sigma^2} -I_* - \delta \right) y ^2 \right]
\label{eq:lem3_eq13}
\end{align}
goes to zero as $y \to +\infty.$ Indeed, $\delta>0$ and the choice of $A$ ensures $\frac{A ^2}{4 \sigma^2} - I_*-\delta > 2\delta > 0$.

We then transfer this concentration to $X_1 = f^*(X_0)$. By \cref{eq:lem2_eq17}, there exists $d > 0$ such that, for $y \ge 1$ and for all $t \in \mathbb R$,
\begin{align*}
\frac{|f ^*(yt)|} {y} & \le 2|t| + \frac{d}{y} \\
& \le 2|t|+d.
\end{align*}

We set
\begin{equation*}
\beta \coloneqq \alpha t_* = \frac{ \alpha^2 \sigma^2}{ 1 + \alpha^2 \sigma^2}.
\end{equation*}

Consequently, there exists a constant $C_\beta>0$ independent of $y$ and $t$ such that
\begin{align*}
\left( \frac{f^*(yt)}{y} - \beta\right)^2 & \le2 \left(\frac{f^*(yt)}{y} \right)^2 + 2 \beta^2 \\
& \le2(2|t| + d)^2 + 2\beta^2 \\
& \le16t^2 + 4d^2 + 2\beta^2 \\
& \le C_\beta(1 + t^2),
\end{align*}
where one may take $C_\beta=\max\{16, 4d^2 + 2\beta ^2\}$.

On the interval $|t - t_*| \le \varepsilon$, our choice $0 < \varepsilon < \frac{t_*}{2}$ gives $0 < t < A$. Hence
\begin{align*}
\left|\frac{f^*(yt)}{y}-\beta\right|
&\le\left|\frac{f^*(yt)}{y}-\alpha t\right|
+\alpha|t-t_*| \\
&\le\sup_{0\le z\le A}
\left|\frac{f^*(yz)}{y}-\alpha z\right|
+\alpha\varepsilon.
\end{align*}

In this proof, let $\mathcal T_y = \{|T_y-t_*| > \varepsilon\}$. Splitting the conditional squared error on $\mathcal T_y$ and its complement, and applying the preceding bounds, we obtain
\begin{align}
&\mathbb E \left[ \left(\frac{X_1}{y}-\beta\right)^2 \mid Y=y \right] \nonumber \\
& = \mathbb E \left[ \left(\frac{X_1}{y} - \beta\right)^2 \mathbbm 1_{\{\mathcal T_y\}} \mid Y=y\right]  + \mathbb E\left[ \left(\frac{X_1}{y}-\beta\right)^2 \mathbbm 1 _{\{\mathcal T_y ^c\}} \mid Y=y\right] \nonumber \\
& \le C_\beta \mathbb E \left[ (1+T_y ^2)\mathbbm 1_{\{\mathcal T_y\}} \mid Y=y\right] + \left( \sup_{0\le t\le A} \left|\frac{f^*(yt)}{y}-\alpha t\right| + \alpha\varepsilon\right)^2 \mathbb P(\mathcal T_y^c\mid Y=y) \nonumber \\
& \le C_\beta\mathbb E \left[ (1+T_y^2)\mathbbm 1_{\{\mathcal T_y\}} \mid Y=y\right]  + \left( \sup_{0\le t\le A} \left|\frac{f^*(yt)}{y}-\alpha t\right| +\alpha\varepsilon\right)^2.
\label{eq:lem3_eq14}
\end{align}

Applying \cref{eq:lem3_eq4} with the fixed endpoint $m = A$ to \cref{eq:lem3_eq13}, taking $y\to+\infty$ gives
\begin{equation*}
\limsup_{y\to+\infty} \mathbb E\left[ \left(\frac{X_1}{y}-\beta\right)^2 \mid Y=y\right] \le\alpha^2\varepsilon^2.
\end{equation*}

This bound holds for every fixed
$0<\varepsilon<\frac{t_*}2$. Since the left-hand side is independent of $\varepsilon$, we have
\begin{equation}
\lim_{y\to+\infty} \mathbb E\left[ \left(\frac{X_1}{y} -\frac{\alpha^2\sigma^2}{1+\alpha^2\sigma^2}\right)^2 \mid Y=y\right]=0.
\label{eq:lem3_eq15}
\end{equation}

We now identify the limiting decoder gain.
Since $g^*(y) = \mathbb E[X_1 \mid Y=y]$, we use conditional Cauchy--Schwarz and \cref{eq:lem3_eq15} to show
\begin{align}
\left|\frac{g^*(y)}{y}-\beta\right|^2 &=\left|\mathbb E\left[ \frac{X_1}{y}-\beta \mid Y=y\right] \right|^2 \nonumber \\
& \le \mathbb E\left[
\left(\frac{X_1}{y}-\beta\right)^2
\mid Y=y\right]
\label{eq:lem3_eq16}
\end{align}
goes to zero when $y \to + \infty$. Thus $\frac{g^*(y)}{y}\to\beta$ as $y\to+\infty$. For every fixed $z \in \mathbb R$, it follows that
\begin{equation*}
\frac{g^*(u+z)}{u} =\frac{g^*(u+z)}{u+z}\frac{u+z}{u}
\end{equation*}
goes to $\beta$ as $u \to + \infty.$ To pass the limit through the expectation defining $\mathcal R_{\mathcal Q^*}$, we use the decoder growth bound in \cref{eq:lem2_eq5}. For $u\ge1$,
\begin{align*}
\left|1-\frac{g^*(u+z)}{u}\right|
&\le1+\frac{|g^*(u+z)|}{u} \\
&\le1+\frac{2|u+z|+C}{u} \\
&\le3+2|z|+C.
\end{align*}

The square of the final expression is integrable against the standard Gaussian density $\varphi(z)$. Therefore, by the dominated convergence theorem,
\begin{align}
\lim_{u\to+\infty}
\frac{\mathcal R_{\mathcal Q^*}(u)}{u^2}
&=\lim_{u\to+\infty}
\int_{z\in\mathbb R}
\left(1-\frac{g^*(u+z)}{u}\right)^2
\varphi(z)\,dz
\nonumber \\
&=\int_{z\in\mathbb R}(1-\beta)^2\varphi(z)\,dz
\nonumber \\
&=(1-\beta)^2
\nonumber \\
&=\frac{1}{(1+\alpha^2\sigma^2)^2}.
\label{eq:lem3_eq17}
\end{align}

Finally, we apply the first controller's best-response condition in \cref{eq:lem2_eq14}. For any fixed $s \ge 0$, choosing the competing action $u = sx$ in \cref{eq:lem2_eq14} gives, for almost every $x$,
\begin{align}
&k^2(f^*(x)-x)^2
+\mathcal R_{\mathcal Q^*}(f^*(x))
\nonumber \\
&\le k^2(s-1)^2x^2
+\mathcal R_{\mathcal Q^*}(sx).
\label{eq:lem3_eq18}
\end{align}

Since $\frac{f^*(x)}{x} \to \alpha>0$, we have $f^*(x)\to+\infty$ as $x \to +\infty$.
By \cref{eq:lem3_eq17},
\begin{equation}
\frac{\mathcal R_{\mathcal Q^*}(f^*(x))}{x^2} =\frac{\mathcal R_{\mathcal Q^*}(f^*(x))}{f^*(x)^2}
\frac{f^*(x)^2}{x^2}
\label{eq:lem3_eq19}
\end{equation}
goes to $\frac{\alpha ^2}{(1+\alpha^2\sigma^2)^2}$ when $x \to + \infty.$ For fixed $s>0$, the same risk limit
\begin{equation*}
\frac{\mathcal R_{\mathcal Q^*}(sx)}{x^2}
=\frac{\mathcal R_{\mathcal Q^*}(sx)}{(sx)^2}s^2
\end{equation*}
goes to $\frac{s^ 2}{(1+\alpha^2\sigma^2)^2}$ as $x \to + \infty.$ For $s = 0$, this limit remains valid because $\mathcal R_{\mathcal Q^*}(0)<\infty$ by \cref{eq:lem2_eq6}. Let $\mathcal G = \mathbb R \setminus \mathcal E$ be the full-measure set of source states on which the best-response inequality in \cref{eq:lem2_eq14} holds for every competing action. Since $\mathcal G$ is unbounded above, we may divide \cref{eq:lem3_eq18} by $x^2$ and let $x \to +\infty$ through $x \in \mathcal G$. The limits above yield
\begin{equation}
k ^2(\alpha-1)^2+\frac{\alpha ^2}{(1+\alpha^2\sigma^2)^2} \le k^2(s-1)^2+\frac{s^2}{(1+\alpha^2\sigma^2)^2} \qquad \forall s \ge0.
\label{eq:lem3_eq20}
\end{equation}

The quadratic function on the right-hand side satisfies
\begin{align}
& k^2(s-1)^2+\frac{s^2}{(1+\alpha^2\sigma^2)^2} \nonumber \\&= \left(k^2+\frac{1}{(1+\alpha^2\sigma^2)^2} \right)s^2-2k^2s+k^2
\nonumber \\
&= \left(k^2+\frac{1}{(1+\alpha^2\sigma^2)^2} \right)
\left(s-\frac{k^2}{k^2+\frac{1}{(1+\alpha^2\sigma^2)^2}}\right)^2
+\frac{k^2\frac{1}{(1+\alpha^2\sigma^2)^2}}{k^2+\frac{1}{(1+\alpha^2\sigma^2)^2}}.
\label{eq:lem3_eq21}
\end{align}

Since the left-hand side of \cref{eq:lem3_eq20} is this same quadratic evaluated at $s=\alpha\ge0$, $\alpha$ must be its unique minimizer. Thus
\begin{gather*}
\alpha =\frac{k^2}{k^2+\frac{1}{(1+\alpha^2\sigma^2)^2}}\in(0,1),
\\
k^2(1-\alpha) = \frac{\alpha}{(1+\alpha^2\sigma^2)^2}.
\end{gather*}

\end{proof}

\paragraph{Consequences of Lemma~\ref{lem_3}.} By \cref{eq:lem3_eq19} and \cref{eq:alpha_condition},
\begin{align}
&\lim_{x\to+\infty}
\frac{k^2(f^*(x)-x)^2
+\mathcal R_{\mathcal Q^*}(f^*(x))}{x^2}
\nonumber \\
&=k^2(1-\alpha)^2+\frac{1}{(1+\alpha^2\sigma^2)^2}\alpha^2
\nonumber \\
&=k^2(1-\alpha)^2+k^2\alpha(1-\alpha)
\nonumber \\
&=k^2(1-\alpha).
\label{eq:lem3_eq22}
\end{align}

Equivalently,
\begin{equation*}
k^2(f^*(x)-x)^2
+\mathcal R_{\mathcal Q^*}(f^*(x))
=k^2(1-\alpha)x^2+o(x^2) 
\end{equation*}
as $x \to + \infty.$

The asymptotic cost above provides a benchmark for testing whether a limiting gain is compatible with global optimality. Lemma~\ref{lem_4} compares it with a control policy that uses a constant encoder output in the tail, yielding a further necessary condition on the limiting gain.

\begin{lemma}
\label{lem_4}

Fix a canonical globally optimal controller pair $(f^*, g^*)$. Suppose that the finite limit $\alpha = \lim_{x \to +\infty} \frac{f^*(x)}x$ exists, then the following inequality holds
\begin{equation}
\sqrt{\alpha^2 + \sigma^{-2}} \ge 1 + \sqrt{1 - \alpha}.
\label{eq:tail-condition}
\end{equation}

Equivalently,
\begin{equation}
\sigma^2 \sqrt{1 - \alpha} (2 - \sqrt{1 - \alpha} ) (1 + \sqrt{1 - \alpha} ) ^2 \le 1.
\label{eq:tail-equivalent}
\end{equation}

\end{lemma}

\begin{proof}

By Lemma~\ref{lem_3}, we have $0 < \alpha <1 $. Suppose, toward a contradiction, that
\begin{equation*}
\sqrt{\alpha ^2 + \sigma ^{-2}} < 1 + \sqrt{1 - \alpha}.
\end{equation*}

Then we can choose a fixed constant $c$ satisfying
\begin{equation}
\max \bigl\{1, \sqrt{ \alpha^2 + \sigma^{-2}} \bigr\} < c < 1 + \sqrt{1 -\alpha}.
\label{eq:lem4_eq1}
\end{equation}

We then choose $\varepsilon > 0$ and $t > \alpha$ such that
\begin{equation}
0 < \varepsilon < c-\sqrt{\alpha^2 + \sigma^{-2}}, \qquad  \alpha<t<\sqrt{(c-\varepsilon)^2-\sigma^{-2}}.
\label{eq:lem4_eq2}
\end{equation}

These choices are attainable because $c > \sqrt{\alpha^2 + \sigma^{-2}}$, which implies
\begin{equation}
(c - 1) ^2 < 1 - \alpha, \qquad \frac{(c - \varepsilon)^2}{1 + \sigma^2 t^2} > \sigma ^{ -2}.
\label{eq:lem4_eq3}
\end{equation}

Indeed, the first inequality follows from $0 < c - 1 < \sqrt{1-\alpha}$, while the second inequality follows by squaring the upper bound on $t$ in \cref{eq:lem4_eq2} and rearranging.

For an arbitrary $T > 0$, define a competing controller pair by
\begin{equation}
\begin{aligned}
f_T(x) & = \begin{cases} f ^*(x), & x < T, \\
cT, & x \ge T \end{cases},\\
g_T(y) & = \begin{cases} cT, &|y -cT| \le \varepsilon T, \\
g^*(y), & |y - cT| > \varepsilon T \end{cases}.
\end{aligned}
\label{eq:lem4_eq4}
\end{equation}

Both controllers are Borel measurable. Moreover, by \cref{eq:lem2_eq5} and \cref{eq:lem2_eq17}, they have at most linear growth for each fixed $T$ and hence finite quadratic cost.

We first determine the probability scale of the modified source tail. For $r > 0$, since $\varphi'(s)=-s\varphi(s)$, integration by parts gives
\begin{align*}
1 - \Phi(r) & = \int_r ^\infty \varphi(s) ds \\
& = \int_r ^\infty \frac{- \varphi'(s)}{s} ds \\
& = \frac{\varphi(r)}{r} - \int_r^ \infty \frac{\varphi(s)}{s^2} ds.
\end{align*}

Because
\begin{equation*}
0 \le \int_r ^\infty \frac{\varphi(s)}{s^2} ds \le \frac{1 - \Phi(r)}{r^2},
\end{equation*}
we obtain the two-sided Gaussian tail bound
\begin{equation}
\frac{r}{1 + r^2} \varphi(r) \le 1-\Phi(r) \le \frac{\varphi(r)}{r}.
\label{eq:lem4_eq5}
\end{equation}

Setting $m_T = \mathbb P(X_0 \ge T)$ and taking $r= \frac{T} \sigma$ yield
\begin{equation*}
m_T=1-\Phi\left(\frac{T}{\sigma}\right),
\end{equation*}
and when $T \to +\infty,$ by \emph{Mills' ratio,} we have
\begin{equation}
\lim_{T \to +\infty} \frac{m_T} {\frac{\sigma}{T\sqrt{2\pi}} \exp \left( -\frac{T^2}{2 \sigma^2} \right)} = 1.
\label{eq:lem4_eq6}
\end{equation}

We also compute the first two truncated moments. Using $\varphi'(s) = -s\varphi(s)$ and integration by parts again, we have
\begin{align*}
\mathbb E \left[X_0 \mathbbm 1 _{\{X_0 \ge T\}} \right] & = \sigma \int_{\frac{T} \sigma} ^\infty s \varphi(s) ds \nonumber \\
& = \sigma \varphi \left( \frac{T}{\sigma} \right),
\end{align*}
and
\begin{align}
\mathbb E \left[X_0 ^2 \mathbbm 1 _{\{X_0 \ge T \}} \right] & = \sigma^2 \int_{\frac{T} \sigma} ^\infty s ^2 \varphi(s) ds \nonumber \\
& = \sigma T \varphi \left(\frac{T}{\sigma} \right) + \sigma^2 m_T.
\label{eq:lem4_eq7}
\end{align}

In particular, \cref{eq:lem4_eq5} gives
\begin{equation*}
1 \le \frac{ \sigma \varphi(\frac{T}{\sigma}) } {Tm_T} \le 1 + \frac{ \sigma^2}{T^2}.
\end{equation*}

Dividing the expressions in \cref{eq:lem4_eq7} by $Tm_T$
and $T^2m_T$, respectively, we conclude that
\begin{equation}
\lim_{T \to +\infty} \frac{ \mathbb E[X_0 \mid X_0 \ge T]}{T} = 1, \qquad \lim_{T \to +\infty}\frac{\mathbb E[X_0^2\mid X_0 \ge T]}{T^2} = 1.
\label{eq:lem4_eq8}
\end{equation}

Next, we bound the probability that the original observation enters the new decoder's window. Throughout this lemma, let $X_1 = f ^*(X_0)$ and $Y = X_1+N$ denote the variables under the original optimal pair. Since $\frac{f^*(x)}{x} \to \alpha$ and $t > \alpha$, we may choose $x_c>0$ such that $f^*(x) \le tx$ for all $x \ge x_c$. The canonical encoder is non-decreasing, so $f ^*(x) \le f^*(x_c)$ for $x \le x_c$. We can choose a constant $A \ge \max \{0, f^*(x_c)\}$ and show
\begin{equation}
f^*(x) \le t\max\{x, 0\} + A \qquad \text{for every } x \in \mathbb R.
\label{eq:lem4_eq9}
\end{equation}

For every $\lambda \ge 0$, \cref{eq:lem4_eq9} and the Gaussian moment-generating formula yield
\begin{align}
\mathbb E [e^{ \lambda X_1} ] & \le e^{A \lambda} \left( \mathbb P(X_0 < 0) + \mathbb E [e^{t\lambda X_0} \mathbbm 1 _{\{X_0 \ge 0 \}} ]\right) \nonumber \\
& \le e^{A \lambda} \left( \frac1 2 + \mathbb E[e^{ t \lambda X_0}] \right) \nonumber \\
& = e^{A \lambda} \left( \frac12 + e^{ \frac{\sigma^2 t^2\lambda^2}2}\right) \nonumber \\
& \le 2 \exp \left(A \lambda + \frac{\sigma^2 t^2 \lambda^2}{2} \right).
\label{eq:lem4_eq10}
\end{align}

Since $\mathbb E[e ^{\lambda N}] = e^{\frac{\lambda^2}2}$, the \emph{exponential Markov inequality} gives, for $\lambda \ge 0$,
\begin{align*}
\mathbb P (Y \ge (c - \varepsilon) T ) & \le e^{- \lambda (c - \varepsilon) T} \mathbb E[e^{\lambda X_1}] \mathbb E[e^{\lambda N}] \\
& \le2 \exp \left( - \lambda ((c-\varepsilon) T - A) + \frac{1 + \sigma^2 t^2}{2} \lambda^2 \right).
\end{align*}

For sufficiently large $T$, the choice
\begin{equation*}
\lambda = \frac{( c - \varepsilon)T- A}{1+\sigma^2 t^2} > 0
\end{equation*}
minimizes the quadratic exponent. Substitution back gives
\begin{equation}
\mathbb P (Y \ge (c-\varepsilon)T ) \le 2 \exp \left( -\frac{((c - \varepsilon)T - A ) ^2} {2(1 + \sigma^2 t^2)} \right).
\label{eq:lem4_eq11}
\end{equation}

Them, differentiating the Gaussian moment-generating function twice gives
\begin{align*}
\mathbb E [N ^2e ^{\lambda N}] & = \frac{d^2}{d \lambda ^2} \mathbb E [e ^{\lambda N}] \\
& = \frac{d^2} {d \lambda^2} e ^{\frac{\lambda^2}2} \\
& = (1 + \lambda ^2) e^{\frac{\lambda^2}2}.
\end{align*}

Applying the same exponential bound, with $\lambda = \frac{(c-\varepsilon)T - A}{1 + \sigma^2 t^2}$, yields
\begin{align}
\mathbb E [N^2\mathbbm 1 _{\{Y \ge (c-\varepsilon)T\}} ] & \le e^{ -\lambda( c- \varepsilon)T} \mathbb E[e^{\lambda X_1}]\mathbb E[N^2e^{\lambda N}] \nonumber\\
& \le 2(1 + \lambda^2) \exp \left(-\frac{((c-\varepsilon)T-A)^2} {2(1+\sigma^2 t^2)}\right).
\label{eq:lem4_eq12}
\end{align}

To compare these estimates with the source-tail mass, we expand the difference of Gaussian exponents
\begin{align}
& - \frac{ ((c-\varepsilon)T - A)^2} {2(1+\sigma^2t^2)}+\frac{T^2}{2\sigma^2} \nonumber \\
& = -\frac12 \left( \frac{(c-\varepsilon)^2}{1+\sigma^2t^2} -\frac1{\sigma^2}\right)T^2 + \frac{(c-\varepsilon)A}{1 + \sigma^2 t^2}T -\frac{A^2}{2(1+\sigma^2 t^2)}.
\label{eq:lem4_eq13}
\end{align}

The coefficient of $-T^2$ is strictly positive by \cref{eq:lem4_eq3}. Consequently, \cref{eq:lem4_eq11}, and \cref{eq:lem4_eq12} imply
\begin{equation}
\mathbb P (Y \ge (c-\varepsilon) T ) = o(m_T), \qquad \mathbb E [N^2 \mathbbm{1}_{\{Y \ge (c - \varepsilon) T \}} ] = o(m_T).
\label{eq:lem4_eq14}
\end{equation}

Indeed, after division by $m_T$, both upper bounds are at most a polynomial in $T$ multiplied by $\exp(-\kappa T^2 + \mathcal O(T))$ for some $\kappa>0$.

We now compare the costs of $(f_T, g_T)$ and $(f^*,g^*)$. On the modified source tail $X_0 \ge T$, the new first-stage cost is
\begin{align}
& k^2 \mathbb E \left[(cT  -X_0) ^2 \mathbbm 1_{\{X_0\ge T\}}\right] \nonumber \\
& = k^2 \left(c ^2 T ^2 m_T -2cT \mathbb E[X_0 \mathbbm 1_{\{X_0 \ge T\}}] +\mathbb E[X_0^2 \mathbbm 1_{\{X_0 \ge T\}}] \right) \nonumber \\
& = k^2m_T T^2 \left( c^2 - 2c \frac{ \mathbb E[X_0 \mid X_0 \ge T]} T + \frac{\mathbb E[X_0^2 \mid X_0 \ge T]}{T^2} \right) \nonumber \\
& = m_T T^2 \big(k^2(c-1)^2 + o(1) \big),
\label{eq:lem4_eq15}
\end{align}
where the last equality uses \cref{eq:lem4_eq8}. On the same source tail, $f_T(X_0) = cT$, so the new observation is $cT + N$. By \cref{eq:lem4_eq4}, the new decoder makes zero error whenever $|N| \le \varepsilon T$. Independence of $X_0$ and $N$
therefore gives
\begin{align}
&\mathbb E [ (f_T(X_0) - g_T(f_T(X_0)+N)  ) ^2 \mathbbm 1_{\{X_0 \ge T\}} ] \nonumber \\
& = m_T\mathbb E [ (cT-g^*(cT + N) )^2 \mathbbm 1_{\{|N|>\varepsilon T\}} ] \nonumber\\
& \le C m_T \mathbb E [ (1+T^2+N^2)\mathbbm 1_{\{|N|>\varepsilon T\}} ] \nonumber\\
& = o(m_TT^2).
\label{eq:lem4_eq16}
\end{align}

Here the bound follows from \cref{eq:lem2_eq5}, since
\begin{align*}
|cT-g^*(cT+N)| &\le cT + 2|cT+N| + C\\
& \le 3cT + 2|N| + C,
\end{align*}
and the final limit holds because
$\mathbb P(|N|>\varepsilon T)$ goes to zero and $\mathbb E[N^2\mathbbm 1_{\{|N|>\varepsilon T\}}]\le1$.

Next, we evaluate the original cost removed from $X_0 \ge T$. By the definition of $\mathcal R_{\mathcal Q^*}$ and independence,
\begin{align*}
& \mathbb E [ (f^*(X_0) - g^*(f^*(X_0)+N) )^2 \mathbbm 1_{\{X_0\ge T\}} ] \\
& = \mathbb E [ \mathcal R_{\mathcal Q^*}(f^*(X_0)) \mathbbm 1_{\{X_0\ge T\}} ].
\end{align*}

The point-wise tail-cost limit in \cref{eq:lem3_eq22} and the tail moments in \cref{eq:lem4_eq8} yield
\begin{align}
& \mathbb E [ (k^2(f^*(X_0) - X_0)^2 + \mathcal R_{\mathcal Q^*} (f^*(X_0)) ) \mathbbm 1_{\{X_0 \ge T\}} ] \nonumber \\
& = m_T T^2 (k^2(1-\alpha)+o(1)).
\label{eq:lem4_eq17}
\end{align}

To justify passage from the point-wise limit to the truncated expectation, note that
\begin{align*}
&\left|\frac{1}{m_TT^2}
\mathbb E\left[
\big(k^2(f^*(X_0)-X_0)^2
+\mathcal R_{\mathcal Q^*}(f^*(X_0))\big)
\mathbbm 1_{\{X_0\ge T\}}\right]
- k^2(1-\alpha)\frac{\mathbb E[X_0^2\mid X_0\ge T]}{T^2}
\right|\\
&\le \sup_{x\ge T}\left|
\frac{k^2(f^*(x)-x)^2
+\mathcal R_{\mathcal Q^*}(f^*(x))}{x^2}
-k^2(1-\alpha)\right|
\frac{\mathbb E[X_0^2\mid X_0\ge T]}{T^2},
\end{align*}
where the supremum tends to zero by \cref{eq:lem3_eq22}, and the final factor tends to one by \cref{eq:lem4_eq8}.

On $X_0<T$, the encoder and the first-stage cost are unchanged. Under the original variables $X_1=f^*(X_0)$ and $Y=X_1 + N$, the second-stage cost differs only on the event $\{|Y-cT|\le\varepsilon T\}$. Since the old squared error is non-negative, we have
\begin{align}
&\mathbb E\left[
\big((X_1-g_T(Y))^2-(X_1-g^*(Y))^2\big)
\mathbbm 1_{\{X_0<T\}}\right]
\nonumber\\
&=\mathbb E\left[
\big((X_1-cT)^2-(X_1-g^*(Y))^2\big)
\mathbbm 1_{\{X_0<T,\ |Y-cT|\le\varepsilon T\}}\right]
\nonumber\\
&\le\mathbb E\left[
(X_1-cT)^2\mathbbm 1_{\{|Y-cT|\le\varepsilon T\}}\right]
\nonumber\\
&=\mathbb E\left[
(Y-cT-N)^2\mathbbm 1_{\{|Y-cT|\le\varepsilon T\}}\right]
\nonumber\\
&\le2\varepsilon^2T^2
\mathbb P\big(Y\ge(c-\varepsilon)T\big)
+2\mathbb E\left[
N^2\mathbbm 1_{\{Y\ge(c-\varepsilon)T\}}\right]
\nonumber\\
&=o(m_TT^2),
\label{eq:lem4_eq18}
\end{align}
where the last step follows from \cref{eq:lem4_eq14}.

Combining \cref{eq:lem4_eq15}, \cref{eq:lem4_eq16}, \cref{eq:lem4_eq17}, and \cref{eq:lem4_eq18}, we obtain
\begin{align}
&J_{k,\sigma}(f_T,g_T)-J_{k,\sigma}(f^*,g^*)
\nonumber\\
&\le m_TT^2\left(
k^2(c-1)^2-k^2(1-\alpha)+o(1)\right).
\label{eq:lem4_eq19}
\end{align}

The coefficient $k^2 ((c-1)^2-(1-\alpha) )$ is strictly negative by \cref{eq:lem4_eq3}. Since $m_T>0$, \cref{eq:lem4_eq19} is strictly negative for all sufficiently large finite $T$. This contradicts the global optimality of $(f^*, g^*)$ and hence this proves \cref{eq:tail-condition}.

Finally, since $0< \alpha <1$, both sides of \cref{eq:tail-condition} are non-negative, and squaring gives
\begin{equation*}
\sigma^{-2}\ge
\bigl(1+\sqrt{1-\alpha}\bigr)^2-\alpha^2.
\end{equation*}

The expression on the right can be rearranged as
\begin{align*}
&\bigl(1+\sqrt{1-\alpha}\bigr)^2-\alpha^2\\
&=\bigl(1+\sqrt{1-\alpha}\bigr)^2
-\bigl((1-\sqrt{1-\alpha})(1+\sqrt{1-\alpha})\bigr)^2\\
&=\bigl(1+\sqrt{1-\alpha}\bigr)^2
\left[1-\bigl(1-\sqrt{1-\alpha}\bigr)^2\right]\\
&=\sqrt{1-\alpha}\bigl(2-\sqrt{1-\alpha}\bigr)
\bigl(1+\sqrt{1-\alpha}\bigr)^2.
\end{align*}

Multiplying by $\sigma^2$ proves \cref{eq:tail-equivalent}.

\end{proof}

Together, Lemmas~\ref{lem_3} and~\ref{lem_4} show that, if an optimal encoder gain converges, its limit must satisfy both the stationary equation and the tail inequality. These conditions restrict the parameter range in which asymptotic linearity is possible. We next establish a matching sufficient condition for the global linear optimality, yielding the exact boundary between linear and non-tame optimality.

\section{Sufficient Condition}

We first state the notations we will use in the following results. For a positive probability density $p$ and a Gaussian density $\varphi_{\sqrt v}$ with variance $v>0$, define the relative entropy and relative Fisher information, respectively by
\begin{gather*}
\mathcal H (p\mid\varphi_{\sqrt v}) \coloneqq \int_{y\in\mathbb R} \left(p(y) \log \frac{p(y)}{\varphi_{\sqrt v}(y)} - p(y) + \varphi_{\sqrt v}(y) \right)dy,
\\
\mathcal I(p\mid\varphi_{\sqrt v}) \coloneqq \int_{y\in\mathbb R}
\left(\frac{p'(y)}{p(y)}+\frac{y}{v}\right)^2p(y) dy.
\end{gather*}

Both integrands are non-negative: for the entropy, this follows from $z\log z-z+1\ge0$ for $z>0$. As in Lemma~\ref{lem_2}, we denote the observation density associated with an output law $\mathcal Q$ by
\begin{equation*}
p_{\mathcal Q}(y)
=\int_{u\in\mathbb R}\varphi(y-u)\,\mathcal Q(du).
\end{equation*}

Gaussian noise smooths differences between an output law and a Gaussian reference. Lemma~\ref{lem_5} quantifies this effect by bounding relative Fisher information and entropy after smoothing in terms of the squared Wasserstein distance between the original laws. This estimate connects the estimation benefit of a non-Gaussian output to its transport cost.

\begin{lemma}
\label{lem_5}

For every positive number $v > 0$ and every $\mathcal Q \in \mathcal W_2(\mathbb R)$, we have
\begin{align}
& \mathcal I(p_{\mathcal Q}\mid\varphi_{\sqrt{1+v}}) + 2\sqrt{\frac{v}{1+v}} \left(1+\sqrt{\frac{v}{1+v}}\right) \mathcal H(p_{\mathcal Q}\mid\varphi_{\sqrt{1+v}}) \nonumber \\
& \le \left(1+\sqrt{\frac{v}{1+v}}\right)^2
W_2 ^2 ( \mathcal Q, \mathcal N(0,v) ).
\label{eq:gaussian_smoothing}
\end{align}

Both quantities on the left-hand side are finite. Moreover, if $\mathcal Q$ is not equivalent to $\mathcal N(0,v)$, then
\begin{align}
\mathcal I(p_{\mathcal Q}\mid\varphi_{\sqrt{1+v}}) &< \left(1+\sqrt{\frac{v}{1+v}}\right)^2 W_2^2 (\mathcal Q, \mathcal N(0,v) ).
\label{eq:gaussian_smoothing_strict}
\end{align}
\end{lemma}

\begin{proof}

We first prove the inequality for a probability law $\mu$ with compact support, under the \emph{Gaussian Ornstein--Uhlenbeck evolution}. For $t \ge 0$, let $\mathcal S_t \mu$ be the law of
\begin{equation}
X_t = e^{-t}X + \sqrt{1 - e ^{-2t}} N,
\label{eq:lem5_eq1}
\end{equation}
where $X \sim \mu,$ and $N \sim \mathcal N(0, 1)$ are independent.

For $t>0$, denote the Lebesgue density of $\mathcal S_t\mu$ by $p_t$ and its density relative to the standard Gaussian law by $h_t(y) = \frac{p_t(y)}{\varphi(y)}$. For convenience, we write
\begin{gather*}
\mathcal H_t \coloneqq \mathcal H(p_t\mid\varphi),
\\
\mathcal I_t \coloneqq \mathcal I(p_t\mid\varphi) =\int_{y\in\mathbb R}\frac{h_t'(y)^2}{h_t(y)} \varphi(y) dy.
\end{gather*}

To justify the differentiations and integrations below, suppose $\operatorname{supp}(\mu) \subseteq[-A, A]$ for some finite $A$. The Gaussian transition density is
\begin{equation}
K_t(x, y) = \frac{1}{\sqrt{2 \pi(1-e^{-2t})}} \exp\left(-\frac{(y-e^{-t}x)^2}{2(1-e^{-2t})} \right),
\label{eq:lem5_eq2}
\end{equation}
so that $p_t(y) = \int K_t(x,y) \mu(dx)$. Fix $0 < s< t < \infty$ and let $u \in [s, t]$. Then, $|e ^{-u}x| \le A$ for $x \in [-A, A]$, while $1-e^{-2u}$ lies in a compact sub-interval of $(0, 1)$. In particular, there exists some positive constants $c_1, c_2, C_1, C_2$ independent of $u \in [s,t]$ and $y \in \mathbb R$ such that
\begin{equation*}
c_1e^{-C_1y^2} \le p_u(y) \le C_2e^{-c_2y^2}.
\end{equation*}

These bounds follow by applying
$(y - e ^{-u}x)^2 \le 2y^2+2(e ^{-u}x)^2$ and $(y - e ^{-u}x) ^2 \ge \frac{y^2}2-(e ^{-u}x)^2$ in the Gaussian kernel, using $|e ^{-u}x| \le A$ and the uniform bounds on its variance. For completeness, direct differentiation of \cref{eq:lem5_eq2} gives
\begin{gather*}
\frac{\partial_y K_u (x,y)}{K_u(x,y)} = -\frac{y - e^{-u}x}{1 - e^{-2u}},
\\
\frac{\partial_y^2K_u(x,y)}{K_u(x,y)}  = \frac{(y-e^{-u}x)^2}{(1 - e^{-2u})^2} -\frac{1}{1-e^{-2u}}.
\end{gather*}

Differentiation with respect to $u$ gives the same type of quadratic bound, since the mean and variance in the kernel have bounded derivatives on $[s, t]$. Upon integrating these formulas against $\mu$, we obtain, uniformly for $u \in [s, t]$,
\begin{equation*}
\left| \frac{p _u'(y)}{p_u(y)}\right| \le C (1 + |y|), \qquad \left| \frac{p_u''(y)}{p_u(y)} \right| + \left| \frac{\partial_u p_u(y)}{p_u(y)} \right| \le C(1 + y^2).
\end{equation*}

The Gaussian lower and upper bounds give $|\log h_u(y)|\le C(1 + y^2)$. Therefore, all differentiations with respect to $u$ and integrations by parts used below are justified by a common polynomial-times-Gaussian integrable bound on $[s,t]$. In particular, $\mathcal H_u$ and $\mathcal I_u$ are finite for every $u>0$.

We next estimate the entropy at a positive time by a quadratic transport cost. Let $\pi(dx, dz)$ be an optimal coupling of $\mu$ and $\mathcal N(0,1)$. Since the Gaussian law is invariant under \cref{eq:lem5_eq1}, we have
\begin{align*}
p_s(y) & = \int K_s(x,y) \pi(dx,dz), \\
\varphi(y) & = \int K_s(z,y) \pi(dx,dz).
\end{align*}

For fixed $y$, the measure
\begin{equation*}
\frac{K_s(z,y)}{\varphi(y)} \pi(dx,dz)
\end{equation*}
is a valid probability measure. Under this measure, the expectation of $\frac{K_s(x, y)}{K_s(z, y)}$ equals $\frac{p_s(y)}{\varphi(y)}$. Since $x \to x \log x$ is convex on $(0, \infty)$, \emph{Jensen's inequality} gives
\begin{align*}
& \frac{p_s(y)} {\varphi(y)} \log \frac{p_s(y)} {\varphi(y)} \\
& \le \int \frac{K_s(x,y)}{K_s(z,y)} \log\frac{K_s(x,y)}{K_s(z,y)} \frac{K_s(z, y)} {\varphi(y)} \pi(dx,dz).
\end{align*}

Multiplying by $\varphi(y)$ and integrating over $y$ yields
\begin{align}
\mathcal H_s
& = \int_{y \in \mathbb R}p_s(y) \log\frac{p_s(y)}{\varphi(y)} dy \nonumber \\
& \le \int_{(x, z) \in \mathbb R^2} \left( \int_{y \in \mathbb R} K_s(x,y) \log\frac{K_s(x,y)}{K_s(z,y)} dy\right) \pi(dx,dz).
\label{eq:lem5_eq3}
\end{align}

We compute the inner integral using \cref{eq:lem5_eq2}
\begin{align*}
\log\frac{K_s(x,y)}{K_s(z,y)} & = \frac{(y-e^{-s}z)^2-(y-e^{-s}x)^2} {2(1-e^{-2s})} \\
& = \frac{2e^{-s}(x-z)y+e^{-2s}(z^2-x^2)} {2(1-e^{-2s})}.
\end{align*}

For a fixed $x$, $K_s(x,\cdot)$ is the Gaussian density with mean \(e^{-s}x\) and variance \(1-e^{-2s}\). Then,
\begin{align*}
&\int_{y\in\mathbb R}K_s(x,y)
\log\frac{K_s(x,y)}{K_s(z,y)}\,dy\\
&=\frac{2e^{-2s}(x-z)x+e^{-2s}(z^2-x^2)}
{2(1-e^{-2s})}\\
&=\frac{e^{-2s}(x-z)^2}{2(1-e^{-2s})}\\
&=\frac{(x-z)^2}{2(e^{2s}-1)}.
\end{align*}

Substituting into \cref{eq:lem5_eq3} and using optimality
of $\pi$ gives
\begin{align}
\mathcal H_s & \le \frac{1}{2(e^{2s} - 1)} \int_{(x, z) \in \mathbb R^2}(x - z)^2 \pi(dx,dz) \nonumber\\
&=\frac{W_2^2(\mu,\mathcal N(0,1))}{2(e^{2s}-1)}.
\label{eq:lem5_eq4}
\end{align}

We then establish contraction of relative Fisher information. For $u \ge 0$, the \emph{Mehler operator} $\mathcal M_u: \mathbb R \to \mathbb R$ acts on a function $h$ by
\begin{equation*}
\mathcal M_uh(y)=\int_{z\in\mathbb R}
 h\left(e^{-u}y+\sqrt{1-e^{-2u}}\,z\right)
\varphi(z)\,dz.
\end{equation*}

The transition kernel satisfies Gaussian detailed balance
\begin{equation*}
K_u(x, y) \varphi(x) = K_u(y, x) \varphi(y).
\end{equation*}

Indeed, both sides have the same constant pre-factor and exponent
\begin{equation*}
-\frac{x^2+y^2-2e^{-u}xy}{2(1-e^{-2u})}.
\end{equation*}

It follows that $h_{s+u} = \mathcal M_u h_s$, and invariance of the standard Gaussian law gives
\begin{equation*}
\int_{y\in\mathbb R} \mathcal M_uh(y)\varphi(y)dy =\int_{y\in\mathbb R}h(y)\varphi(y)dy.
\end{equation*}

Differentiating the Mehler formula yields
\begin{equation*}
\frac{d}{dy} \mathcal M_uh_s(y) = e^{-u} \mathcal M_uh_s'(y).
\end{equation*}

By Cauchy--Schwarz with respect to the Gaussian kernel in $\mathcal M_u$, we obtain
\begin{align*}
\bigl(\mathcal M_uh_s'(y)\bigr)^2 & = \left(\mathcal M_u \left( \frac{h_s'}{\sqrt{h_s}}\sqrt{h_s} \right )(y) \right) ^2 \\
& \le \mathcal M_u\left (\frac{(h_s')^2}{h_s}\right ) (y) \mathcal M_u h_s(y).
\end{align*}

Since $\mathcal M_uh_s>0$, division and integration against $\varphi(y) dy$ give
\begin{align}
\mathcal I_{s+u} &=e^{-2u}\int_{y\in\mathbb R} \frac{\bigl(\mathcal M_uh_s'(y) \bigr)^2}{\mathcal M_uh_s(y)} \varphi(y) dy \nonumber \\
& \le e^{-2u}\int_{y \in \mathbb R} \mathcal M_u \left( \frac{(h_s')^2}{h_s} \right)(y) \varphi(y) dy \nonumber \\
& = e^{-2u}\int_{y \in \mathbb R} \frac{h_s'(y)^2}{h_s(y)} \varphi(y) dy \nonumber\\
&=e^{-2u} \mathcal I_s.
\label{eq:lem5_eq5}
\end{align}

Next, we compute entropy dissipation. The density $p_t$ solves $\partial_tp_t = \partial_y^2p_t + \partial_y(yp_t)$, which follows by differentiating the Gaussian mixture in \cref{eq:lem5_eq2}. Since $p_t=h_t \varphi$ and $\varphi'(y)=-y\varphi(y)$,
\begin{align*}
\partial_t(h_t\varphi) & = \partial_y^2(h_t\varphi) +\partial_y(yh_t\varphi) \\
& = \partial_y (h_t'\varphi ).
\end{align*}

Therefore $\partial_th_t=h_t''-yh_t'$. By differentiation under the integral and integration by parts,
\begin{align}
\frac{d \mathcal H_t}{dt} & = \int_{y\in\mathbb R}(1+\log h_t(y)) \partial_th_t(y)\varphi(y) dy \nonumber\\
& = \int_{y\in\mathbb R}(1+\log h_t(y)) \partial_y (h_t'(y)\varphi(y) ) dy \nonumber\\
&=-\int_{y\in\mathbb R} \frac{h_t'(y)^2}{h_t(y)}\varphi(y) dy \nonumber\\
& = - \mathcal I_t.
\label{eq:lem5_eq6}
\end{align}

Integrating \cref{eq:lem5_eq6} between $s$ and $t$, where
$0<s<t$, gives
\begin{equation*}
\mathcal H_s- \mathcal H_t = \int_s^t \mathcal I_u du.
\end{equation*}

To obtain a lower bound on this integral, apply \cref{eq:lem5_eq5} with starting time $u$ and elapsed time $t-u$.

Since $\mathcal I_t \le e^{-2(t-u)} \mathcal I_u$, we have $\mathcal I_u \ge e^{2(t-u)} \mathcal I_t$. Hence
\begin{align}
\mathcal H_s - \mathcal H_t & = \int_s ^t \mathcal I_u du \nonumber\\
& \ge \mathcal I_t \int_s^t e^{2(t-u)}du \nonumber\\
&=  \mathcal I_t \frac{e^{2(t-s)}-1}{2}.
\label{eq:lem5_eq7}
\end{align}

Combining \cref{eq:lem5_eq4} and \cref{eq:lem5_eq7} gives, for every $0< s <t$,
\begin{equation*}
\mathcal H_t + \frac{e^{2(t-s)}-1}{2} \mathcal I_t
\le\frac{W_2^2(\mu,\mathcal N(0,1))}{2(e^{2s}-1)}.
\end{equation*}

Choosing $s= \frac t{2}$ makes the two exponential factors $e^{2s}-1$ and $e^{2(t-s)}-1$ equal. Therefore
\begin{align*}
\mathcal H_t + \frac{e^t-1}{2} \mathcal I_t
& \le\frac{W_2^2(\mu,\mathcal N(0,1))}{2(e^t-1)}, \\
\mathcal I_t + \frac{2}{e^t-1} \mathcal H_t
&\le\frac{W_2^2(\mu,\mathcal N(0,1))}{(e^t-1)^2}.
\end{align*}

Thus we have established the Gaussian evolution inequality
\begin{equation}
\mathcal I_t+ \frac{2}{e^t-1} \mathcal H_t
\le\frac{W_2^2(\mu,\mathcal N(0,1))}{(e^t-1)^2}.
\label{eq:lem5_eq8}
\end{equation}

We now transfer the preceding inequality to the Gaussian observation channel $Y=X_1+N$, where $X_1\sim\mathcal Q$ and $N \sim \mathcal N(0,1)$ are independent. For any fixed $v > 0$, our goal is to represent the normalized observation
$\frac{Y}{\sqrt{1+v}}$ through the Ornstein--Uhlenbeck evolution in \cref{eq:lem5_eq1}.

We first suppose that $\mathcal Q$ has compact support. Let $X_1\sim\mathcal Q$ and define the auxiliary probability measure $\mu$ as the distribution of $\frac{X_1}{\sqrt v}.$ We then choose the evolution time $t > 0$ such that
\begin{equation*}
e^{-2t}=\frac{v}{1+v},
\end{equation*}
or equivalently,
\begin{equation*}
t=\frac{1}{2}\log\left(\frac{1+v}{v}\right).
\end{equation*}

Substituting these choices into
\cref{eq:lem5_eq1}, we obtain
\begin{align}
e^{-t}\frac{X_1}{\sqrt v}
+\sqrt{1-e^{-2t}}N
&=\sqrt{\frac{v}{1+v}}
\frac{X_1}{\sqrt v}
+\sqrt{\frac{1}{1+v}}N
\nonumber\\
&=\frac{X_1+N}{\sqrt{1+v}}
\nonumber\\
&=\frac{Y}{\sqrt{1+v}}.
\label{eq:lem5_eqX}
\end{align}

Consequently, $\mathcal S_t\mu$ is the distribution of $\frac{Y}{\sqrt{1+v}},$ with Lebesgue density
\begin{equation*}
p_t(y)=\sqrt{1+v}
p_{\mathcal Q}(\sqrt{1+v} y).
\end{equation*}

Writing $z=\sqrt{1+v}y$, we have
\begin{align*}
\frac{p_t(y)}{\varphi(y)}
&=\frac{\sqrt{1+v}\,p_{\mathcal Q}(z)}{\varphi(y)}\\
&=\frac{p_{\mathcal Q}(z)}{\varphi_{\sqrt{1+v}}(z)},
\end{align*}
and
\begin{equation*}
\frac{d}{dy}\log\frac{p_t(y)}{\varphi(y)} =\sqrt{1+v}\left(
\frac{p_{\mathcal Q}'(z)}{p_{\mathcal Q}(z)}
+\frac{z}{1+v}\right).
\end{equation*}

Since $p_t(y)dy = p_{\mathcal Q}(z)dz$, the relative entropy and Fisher information transform as
\begin{gather}
\mathcal H_t = \mathcal H (p_{\mathcal Q} \mid \varphi_{\sqrt{1+v}}), \nonumber \\
\mathcal I_t = (1+v)\mathcal I
(p_{\mathcal Q} \mid \varphi_{\sqrt{1+v}}).
\label{eq:lem5_eq9} 
\end{gather}

Moreover, simultaneous scaling of the coordinates of any coupling by $\frac{1}{\sqrt v}$ proves
\begin{equation*}
W_2^2(\mu,\mathcal N(0,1)) =\frac{1}{v}W_2^2(\mathcal Q,\mathcal N(0,v)).
\end{equation*}

Substituting these identities into \cref{eq:lem5_eq8} and dividing by $1 + v$ gives
\begin{align}
&\mathcal I(p_{\mathcal Q}\mid\varphi_{\sqrt{1+v}})
+\frac{2}{(1+v)(e^t-1)}
\mathcal H(p_{\mathcal Q}\mid\varphi_{\sqrt{1+v}})
\nonumber\\
&\le\frac{1}{v(1+v)(e^t-1)^2}
W_2^2(\mathcal Q,\mathcal N(0,v)).
\label{eq:lem5_eq10}
\end{align}

The coefficients in \cref{eq:lem5_eq10} must be evaluated
with the selected value
$e^t=\sqrt{ \frac{1+v}v}$. For the transport coefficient,
\begin{align*}
\frac{1}{v(1+v)(e^t-1)^2}
&=\frac{1}{(1+v)(\sqrt{1+v}-\sqrt v)^2}\\
&=\frac{(\sqrt{1+v}+\sqrt v)^2}{1+v}\\
&=\left(1+\sqrt{\frac{v}{1+v}}\right)^2,
\end{align*}
where we used
$(\sqrt{1+v}-\sqrt v)(\sqrt{1+v}+\sqrt v)=1$.
Similarly, the entropy coefficient satisfies
\begin{align*}
\frac{2}{(1+v)(e^t-1)}
&=\frac{2\sqrt v}{(1+v)(\sqrt{1+v}-\sqrt v)}\\
&=\frac{2\sqrt v(\sqrt{1+v}+\sqrt v)}{1+v}\\
&=2\sqrt{\frac{v}{1+v}}
\left(1+\sqrt{\frac{v}{1+v}}\right).
\end{align*}

Hence \cref{eq:gaussian_smoothing} holds whenever
$\mathcal Q$ has compact support.

Let $X_1\sim\mathcal Q$ and define the clipped random variable and its law by
\begin{equation*}
X_{1n}=\max\{-n,\min\{X_1,n\}\}.
\end{equation*}

Correspondingly, $\mathcal Q_n$ is the distribution of $X_{1n}.$ Because $\mathcal Q$ has finite second moment, the coupling $(X_{1n},X_1)$ gives
\begin{align}
W_2^2(\mathcal Q_n,\mathcal Q)
&\le\mathbb E[(X_{1n}-X_1)^2]
\nonumber\\
&\le\mathbb E[X_1^2 \mathbbm 1 _{\{ |X_1|>n \}}].
\label{eq:lem5_eq11}
\end{align}

The right-hand side converges to zero as $n \to \infty$. By the triangle inequality for $W_2$, this implies
\begin{equation*}
W_2(\mathcal Q_n,\mathcal N(0,v))
\to W_2(\mathcal Q,\mathcal N(0,v)).
\end{equation*}

For every fixed $y$, the functions $u \to \varphi(y-u)$ and $u \to \varphi'(y-u)$ are bounded and continuous. Since $X_{1n} \to X_1$ almost surely, dominated convergence theorem gives
\begin{equation*}
p_{\mathcal Q_n}(y) \to p_{\mathcal Q}(y)>0, \qquad p_{\mathcal Q_n}'(y)\to p_{\mathcal Q}'(y).
\end{equation*}

Consequently, the two non-negative integrands $\frac{\left(p_{\mathcal Q_n}'(y) + \frac{y}{1+v}p_{\mathcal Q_n}(y)\right)^2} {p_{\mathcal Q_n}(y)}$ and  $ p_{\mathcal Q_n}(y) \log\frac{p_{\mathcal Q_n}(y)}{\varphi_{\sqrt{1+v}}(y)} -p_{\mathcal Q_n}(y) + \varphi_{\sqrt{1+v}}(y)$ converge point-wise to their corresponding expressions with $\mathcal Q$ in place of $\mathcal Q_n$. \emph{Fatou's lemma}, applied to their sum with the positive entropy coefficient in \cref{eq:gaussian_smoothing}, yields
\begin{align}
&\mathcal I(p_{\mathcal Q}\mid\varphi_{\sqrt{1+v}})
+2\sqrt{\frac{v}{1+v}}
\left(1+\sqrt{\frac{v}{1+v}}\right)
\mathcal H(p_{\mathcal Q}\mid\varphi_{\sqrt{1+v}})
\nonumber\\
&\le\liminf_{n\to\infty}\Bigg(
\mathcal I(p_{\mathcal Q_n}\mid\varphi_{\sqrt{1+v}})
+2\sqrt{\frac{v}{1+v}}
\left(1+\sqrt{\frac{v}{1+v}}\right)
\mathcal H(p_{\mathcal Q_n}\mid\varphi_{\sqrt{1+v}})
\Bigg)
\nonumber\\
&\le\left(1+\sqrt{\frac{v}{1+v}}\right)^2
W_2^2(\mathcal Q,\mathcal N(0,v)).
\label{eq:lem5_eq12}
\end{align}

The final quantity is finite, which also proves the claimed finiteness of both terms on the left.

To establish strictness, recall that $z\log z-z+1\ge0$ for $z>0$, with equality only at $z=1$. Thus
\begin{equation*}
\mathcal H(p_{\mathcal Q}\mid\varphi_{\sqrt{1+v}})=0
\end{equation*}
holds only if $p_{\mathcal Q}(y)=\varphi_{\sqrt{1+v}}(y)$
for almost every $y$. Taking characteristic
functions gives
\begin{align*}
\mathbb E [e^{i\xi X_1}]e^{- \frac{\xi^2}2}
&=\mathbb E[e^{i\xi Y}]\\
&= e^{-\frac{(1+v)\xi^2}2},\\
\mathbb E[e^{i\xi X_1}] &=e^{-\frac{v\xi^2}2},
\end{align*}
for every $\xi\in\mathbb R$.
By uniqueness of characteristic functions,
$\mathcal Q=\mathcal N(0,v)$.
Hence, for $\mathcal Q\ne\mathcal N(0,v)$,
the entropy remainder in \cref{eq:gaussian_smoothing}
is strictly positive, yielding
\cref{eq:gaussian_smoothing_strict}.
\end{proof}

Under the stationary condition, Lemma~\ref{lem_6} expresses the cost difference from a Gaussian candidate as a Wasserstein term minus relative Fisher information. The bound in Lemma~\ref{lem_5} then yields a sufficient condition for the Gaussian candidate to be the unique global optimum.

\begin{lemma}
\label{lem_6}

Suppose $0 < \alpha < 1$ satisfies \cref{eq:alpha_condition}. Let $v = \sigma^2 \alpha^2$. We have
\begin{align}
&J_{k,\sigma}(\mathcal Q)-J_{k,\sigma}(\mathcal N(0,v)) \nonumber\\
&=\left(k^2+\frac{1}{(1+v)^2}\right)
W_2^2 (\mathcal Q,\mathcal N(0,v) )
-\mathcal I(p_{\mathcal Q} \mid \varphi_{\sqrt{1+v}})
\label{eq:gaussian_comparison}
\end{align}
for every $\mathcal Q \in \mathcal W_2(\mathbb R).$ In particular, suppose
\begin{equation}
k^2+\frac{1}{(1+v)^2} \ge\left(1+\sqrt{\frac{v}{1+v}}\right)^2,
\label{eq:gaussian_coefficient}
\end{equation}
then the unique optimal controller pair, up to null sets, is
\begin{equation}
f^*(x)=\alpha x,
\qquad
g^*(y)=\frac{\alpha^2\sigma^2}{1+\alpha^2\sigma^2}\,y.
\label{eq:gaussian_optimal_pair}
\end{equation}

\end{lemma}

\begin{proof}

Fix an arbitrary $\mathcal Q\in\mathcal W_2(\mathbb R)$. We first compare the MMSE decoder $g_{\mathcal Q}$ with the linear decoder $y \to\frac{v}{1+v}y$. By differentiating the Gaussian convolution, we have
\begin{align*}
p_{\mathcal Q}'(y)
&=\int_{u\in\mathbb R}\varphi'(y-u)\,\mathcal Q(du)\\
&=\int_{u\in\mathbb R}(u-y)\varphi(y-u)\,\mathcal Q(du)\\
&=\int_{u\in\mathbb R}u\varphi(y-u)\,\mathcal Q(du)
-y p_{\mathcal Q}(y).
\end{align*}

The differentiation is justified because $\varphi'$ is
bounded and continuous. Since $p_{\mathcal Q}(y)>0$ and
$g_{\mathcal Q}(y)$ is the corresponding conditional mean,
\begin{align}
g_{\mathcal Q}(y)
&=\frac{\displaystyle\int_{u\in\mathbb R}
 u\varphi(y-u)\,\mathcal Q(du)}{p_{\mathcal Q}(y)}
\nonumber\\
&=y+\frac{p_{\mathcal Q}'(y)}{p_{\mathcal Q}(y)}.
\label{eq:lem6_eq1}
\end{align}

Consequently,
\begin{align*}
g_{\mathcal Q}(y)-\frac{v}{1+v}y
&=\left(1-\frac{v}{1+v}\right)y
+\frac{p_{\mathcal Q}'(y)}{p_{\mathcal Q}(y)}\\
&=\frac{p_{\mathcal Q}'(y)}{p_{\mathcal Q}(y)}
+\frac{y}{1+v}.
\end{align*}

Integrating the square against the observation density,
and using the definition of relative Fisher information,
we obtain the exact identity
\begin{align}
& \mathbb E [ (g_{\mathcal Q}(Y)-\frac{v}{1+v}Y )^2 ] \nonumber \\
& = \int_{y \in \mathbb R} \left( \frac{p_{\mathcal Q}'(y)}{p_{\mathcal Q}(y)} + \frac{y}{1+v}\right)^2p_{\mathcal Q}(y) dy \nonumber\\
& = \mathcal I(p_{\mathcal Q}\mid\varphi_{\sqrt{1+v}}).
\label{eq:lem6_eq2}
\end{align}

We next use the orthogonality of conditional expectation.
Since $g_{\mathcal Q}(Y) = \mathbb E[X_1\mid Y]$,
\begin{equation*}
\mathbb E [ (X_1-g_{\mathcal Q}(Y)) (g_{\mathcal Q}(Y)-\frac{v}{1+v}Y ) ]=0.
\end{equation*}

Expanding the square and applying \cref{eq:lem6_eq2},
\begin{align}
& \mathbb E [ (X_1-\frac{v}{1+v}Y )^2 ] \nonumber \\
& = \mathbb E[(X_1-g_{\mathcal Q}(Y))^2] + \mathbb E [ (g_{\mathcal Q}(Y) - \frac{v}{1+v}Y )^2 ] \nonumber \\
& = \textit{mmse}(\mathcal Q,1) + \mathcal I(p_{\mathcal Q}\mid\varphi_{\sqrt{1+v}}).
\label{eq:lem6_eq3}
\end{align}

On the other hand, substituting $Y = X_1+N$ gives
\begin{align*}
\mathbb E [ (X_1-\frac{v}{1+v}Y )^2 ]
& = \mathbb E [ (\frac{X_1}{1+v}-\frac{vN}{1+v} )^2 ]\\
& = \frac{\mathbb E[X_1^2]}{(1+v)^2} -\frac{2v}{(1+v)^2}\mathbb E[X_1N] +\frac{v^2}{(1+v)^2}\mathbb E[N^2]\\
& = \frac{\mathbb E[X_1^2]+v^2}{(1+v)^2}.
\end{align*}

Here $\mathbb E[X_1N] = \mathbb E[X_1]\mathbb E[N] = 0$ by independence, and $\mathbb E[N^2]=1$. Combining this calculation with \cref{eq:lem6_eq3}, we obtain
\begin{equation}
\textit{mmse}(\mathcal Q,1) = \frac{\mathbb E[X_1^2]+v^2}{(1+v)^2} -\mathcal I(p_{\mathcal Q}\mid\varphi_{\sqrt{1+v}}).
\label{eq:lem6_eq4}
\end{equation}

All quantities in this identity are finite: the first two moments are finite because $\mathcal Q\in\mathcal W_2(\mathbb R)$, and we also have $\mathbb E[g_{\mathcal Q}(Y) ^2] < \infty$ by conditional Jensen's inequality.

We now combine \cref{eq:lem6_eq4} with the optimal transport reduction in Lemma~\ref{lem_1}. In one dimension, the quadratic Wasserstein distance is obtained by coupling corresponding quantiles. Let $f_{\mathcal Q}$ be the non-decreasing quadratic optimal transport map from $\mathcal P$ to $\mathcal Q$. Therefore,
\begin{equation*}
W_2^2(\mathcal P,\mathcal Q) = \int_{x \in \mathbb R} (x-f_{\mathcal Q}(x))^2 \varphi_\sigma(x) dx,
\end{equation*}
and
\begin{equation*}
\mathbb E[X_1^2] =\int_{x\in\mathbb R} f_{\mathcal Q}(x)^2\varphi_\sigma(x) dx.
\end{equation*}

Moreover, the map $x \to \alpha x$ is increasing because $\alpha>0$, and it transports from $\mathcal P$ to $\mathcal N(0,v)$ because $v = \alpha^2\sigma^2$. Thus $f_{\mathcal Q}(X_0)$ and $\alpha X_0$ form a non-decreasing coupling of $\mathcal Q$ and $\mathcal N(0,v)$. This coupling is optimal for the one-dimensional quadratic transport cost: if $u_1\le u_2$ and $z_1\le z_2$, then
\begin{align*}
&(u_1-z_2)^2+(u_2-z_1)^2\\
& -(u_1-z_1)^2-(u_2-z_2)^2\\
& =2(u_2-u_1)(z_2-z_1)\ge0.
\end{align*}

Hence the rearrangement principle used in Lemma~\ref{lem_1} yields
\begin{equation*}
W_2^2\bigl(\mathcal Q,\mathcal N(0,v)\bigr)
=\int_{x\in\mathbb R}
(f_{\mathcal Q}(x)-\alpha x)^2
\varphi_\sigma(x)\,dx.
\end{equation*}

By the stationary condition in \cref{eq:alpha_condition},
\begin{gather}
k^2 (1- \alpha ) = \frac{\alpha}{(1 + \sigma ^2 \alpha^2) ^2}, \nonumber \\
k^2\alpha+\frac{\alpha}{(1+v)^2} =k^2, \nonumber \\
 (k^2 + \frac{1}{(1+v)^2} )\alpha = k^2.
\label{eq:lem6_eq5}
\end{gather}

Using the transport identities above, we expand
\begin{align}
& k^2 W_2^2(\mathcal P, \mathcal Q) + \frac{\mathbb E[X_1^2]}{(1+v)^2} \nonumber \\
& = \int_{x \in \mathbb R} \big( k^2(x-f_{\mathcal Q}(x))^2 +\frac{f_{\mathcal Q}(x)^2}{(1+v)^2} \big) \varphi_\sigma(x) dx \nonumber \\
& = \int_{x \in \mathbb R} \big( (k^2+\frac{1}{(1 + v)^2} ) f_{\mathcal Q}(x)^2 -2k^2x f_{\mathcal Q}(x) +k^2x^2 \big) \varphi_\sigma(x) dx.
\label{eq:lem6_eq6}
\end{align}

By \cref{eq:lem6_eq5}, the cross term satisfies
\begin{equation*}
2k^2 x f_{\mathcal Q}(x) =2 \big(k^2+\frac{1}{(1+v)^2} \big) \alpha x f_{\mathcal Q}(x).
\end{equation*}

Consequently, completing the square in \cref{eq:lem6_eq6}, we obtain
\begin{align}
k ^2 W_2^2(\mathcal P, \mathcal Q) + \frac{\mathbb E[X_1^2]}{(1+v)^2} & = \int_{x \in \mathbb R} \Big( \big(k^2 + \frac{1}{(1 + v)^2}\big) (f_{\mathcal Q}(x)-\alpha x)^2 \nonumber\\
& + \big(k^2- \big(k^2 + \frac{1}{(1+v)^2} \big) \alpha^2 \big) x^2 \Big) \varphi_ \sigma(x) dx \nonumber \\
& = \big(k^2 + \frac{1}{(1+v) ^2} \big) W_2 ^2 (\mathcal Q, \mathcal N(0, v) ) \nonumber \\
& + k ^2 \sigma^2 - \big(k^2+\frac{1}{(1+v)^2}\big)v.
\label{eq:lem6_eq7}
\end{align}

Here the final equality follows from $\int_{x \in \mathbb R}x^2 \varphi_ \sigma(x) dx = \sigma^2$ and $\alpha^2\sigma^2 = v$. This identity holds for every $\mathcal Q \in \mathcal W_2(\mathbb R)$, including non-Gaussian laws with atoms.

By the optimal transport reduction in Lemma~\ref{lem_1}, substituting \cref{eq:lem6_eq4} and \cref{eq:lem6_eq7}
gives
\begin{align}
J_{k,\sigma}(\mathcal Q) & = k^2 W_2^2(\mathcal P, \mathcal Q) + \textit{mmse}(\mathcal Q, 1) \nonumber \\
& = \big( k^2 + \frac{1}{(1 + v)^2} \big) W_2 ^2 (\mathcal Q,\mathcal N(0,v) ) \nonumber\\
& +k^2\sigma^2 -\big(k^2+\frac{1}{(1+v)^2}\big)v +\frac{v^2}{(1+v)^2} -\mathcal I(p_{\mathcal Q}\mid\varphi_{\sqrt{1+v}}).
\label{eq:lem6_eq8}
\end{align}

At $\mathcal Q=\mathcal N(0,v)$, the Wasserstein term vanishes. The observation density is exactly $\varphi_{\sqrt{1+v}}$, so the relative Fisher information also vanishes. Subtracting the identity from \cref{eq:lem6_eq8} proves \cref{eq:gaussian_comparison}.

We now apply Lemma~\ref{lem_5}. Rearranging the inequality
in \cref{eq:gaussian_smoothing} gives
\begin{align*}
-\mathcal I(p_{\mathcal Q}\mid\varphi_{\sqrt{1+v}})
& \ge - \left(1+\sqrt{\frac{v}{1+v}}\right)^2
W_2^2(\mathcal Q,\mathcal N(0,v))\\
& + 2\sqrt{\frac{v}{1+v}}
\left(1+\sqrt{\frac{v}{1+v}}\right)
\mathcal H(p_{\mathcal Q}\mid\varphi_{\sqrt{1+v}}).
\end{align*}

Substituting this bound into \cref{eq:gaussian_comparison},
we obtain
\begin{align}
&J_{k,\sigma}(\mathcal Q)-J_{k,\sigma}(\mathcal N(0,v))
\nonumber\\
&\ge\left(k^2+\frac{1}{(1+v)^2}
-\left(1+\sqrt{\frac{v}{1+v}}\right)^2\right) W_2^2 (\mathcal Q,\mathcal N(0,v) ) \nonumber\\
& +2\sqrt{\frac{v}{1+v}}
\left(1+\sqrt{\frac{v}{1+v}}\right)
\mathcal H(p_{\mathcal Q}\mid\varphi_{\sqrt{1+v}}).
\label{eq:gaussian_cost_bound}
\end{align}

Suppose now that \cref{eq:gaussian_coefficient} holds.
The Wasserstein coefficient in \cref{eq:gaussian_cost_bound}
is non-negative, while the entropy coefficient is strictly positive because $v>0$. Therefore
\begin{equation*}
J_{k, \sigma}(\mathcal Q) \ge J_{k,\sigma}(\mathcal N(0, v))
\end{equation*}
for every $\mathcal Q\in\mathcal W_2(\mathbb R)$. Moreover, if $\mathcal Q\ne\mathcal N(0,v)$, Lemma~\ref{lem_5} shows that the entropy is strictly positive. Hence
\begin{equation*}
J_{k,\sigma}(\mathcal Q) >J_{k,\sigma}(\mathcal N(0,v)), \qquad \mathcal Q\ne\mathcal N(0,v).
\end{equation*}

This proves uniqueness of the minimizing law even when equality holds in \cref{eq:gaussian_coefficient}.

It remains to recover the controllers. Since $X_0 \sim \mathcal N(0,\sigma^2)$ and the unique optimal output law is $\mathcal N(0,v)$, the unique non-decreasing quadratic transport is
\begin{equation*}
f^*(x) = \frac{\sqrt v}{\sigma}x = \alpha x.
\end{equation*}

For this encoder, $Y=\alpha X_0+N$ is Gaussian, with variance $1+v$ and covariance $\operatorname{Cov}(X_1,Y)=v$. Its conditional mean is
\begin{align*}
g^*(y) & =\mathbb E[X_1\mid Y=y] \\
& = \frac{\operatorname{Cov}(X_1,Y)}{\operatorname{Var}(Y)}y \\
& = \frac{v}{1+v}y \\
& =\frac{\alpha^2 \sigma^2}{1 + \alpha^2 \sigma^2}y.
\end{align*}

The equality conditions for conditional expectation and one-dimensional quadratic transport give uniqueness of both controllers up to their corresponding null sets. Since the canonical encoder is non-decreasing and right-continuous, and agrees almost everywhere with the continuous function $x\to\alpha x$, it agrees with this function everywhere. This completes the proof.
\end{proof}

Under the stationary condition, the necessary tail inequality in Lemma~\ref{lem_4} is equivalent to the sufficient condition for global linear optimality in Lemma~\ref{lem_6}. This equivalence yields the exact boundary between linear and nonlinear optimality, expressed by the critical curve defined below.

\begin{definition}[critical curve]
\label{def:curve}

For any $\sigma \in( \frac{1}{2}, \infty)$, let $q_\sigma \in (0,1)$ denote the unique solution of
\begin{equation}
\sigma^2 q_\sigma(2-q_\sigma)(1+q_\sigma)^2 = 1.
\label{eq:critical_q}
\end{equation}

Correspondingly, we define the critical parameter by
\begin{equation}
k_c(\sigma)= \begin{cases}
0, & 0 < \sigma \le \frac12, \\
(2 - q_\sigma) \sqrt{1 - q_\sigma^2}, & \sigma > \frac12 \end{cases}.
\label{eq:critical_k}
\end{equation}
    
\end{definition}

Below, Theorem~\ref{thm_1} characterizes globally optimal controllers across this critical curve.

\begin{theorem}
\label{thm_1}

For arbitrary positive parameters $k$ and $\sigma$, the globally optimal controllers $(f ^*, g ^*)$ exhibit the following dichotomy: suppose $k \in [k_c ( \sigma), + \infty)$, then the globally optimal controller pair is essentially unique and linear, and admits the following exact representation
\begin{equation}
f ^*(x) = \alpha x, \qquad g ^*(y) = \frac{ \alpha ^2 \sigma ^2} {1 + \alpha^2 \sigma^2} y,
\label{eq:linear_optimal_pair}
\end{equation}
where $\alpha \in (0,1)$ is the unique solution in Lemma~\ref{lem_3}.

In contrast, suppose $k \in (0, k_c ( \sigma))$, then any canonical globally optimal encoder $f ^*$ fails to be asymptotically linear on either $x \to + \infty$ or $x \to - \infty$. More precisely,
\begin{equation}
0 < \liminf_{x \to \infty} \frac{f ^*(x)} {x} < \limsup_{x \to \infty} \frac{f ^*(x)} {x} < \infty.
\label{eq:nonconvergent_gain}
\end{equation}

\end{theorem}

\begin{proof}
We first establish that the necessary tail inequality and the sufficient condition for Gaussian optimality impose exactly the same scalar restriction whenever the stationary equation holds. Fix $\alpha \in (0, 1)$ satisfying \cref{eq:alpha_condition}, and write $v = \sigma^2 \alpha^2 > 0$. The stationary condition gives
\begin{align}
k ^2 + \frac{1}{(1 + v)^2} & = \frac{ \alpha} {(1 - \alpha)(1 + v)^2} + \frac{1}{(1 + v)^2} \nonumber \\
& =\frac{\alpha + 1 - \alpha}{(1 - \alpha)(1 + v)^2} \nonumber \\
& = \frac{1}{(1 - \alpha)(1 + v)^2}.
\label{eq:thm1_eq1}
\end{align}

In view of \cref{eq:thm1_eq1}, the coefficient condition in \cref{eq:gaussian_coefficient} is equivalent to
\begin{align}
& \frac{1} {(1 - \alpha)(1 + v)^2} \ge \left( 1 + \sqrt{\frac{v}{1 + v}} \right) ^2 \nonumber \\
& \iff \frac{1} {\sqrt{1 - \alpha}(1 + v)} \ge 1 + \sqrt{\frac{v}{1 + v}} \nonumber \\
& \iff \sqrt{1 - \alpha} \le \frac{1}{(1 + v) \left(1  +\sqrt{\frac{v}{1 + v}} \right)} \nonumber \\
& \iff \sqrt{1 - \alpha} \le1 - \sqrt{ \frac{v}{1 + v}} \nonumber \\
& \iff \frac{ \alpha \sigma} {\sqrt{1 + \alpha^2 \sigma^2}} \le 1 - \sqrt{1 - \alpha} \nonumber \\
& \iff \sqrt{\alpha^2 + \sigma^{-2}} \ge \frac{\alpha}{1 - \sqrt{1 - \alpha}} \nonumber \\
& \iff \sqrt{ \alpha^2 + \sigma^{-2}} \ge 1 + \sqrt{1 - \alpha}.
\label{eq:thm1_eq2}
\end{align}

Here the third equivalence uses
\begin{align*}
& (1+v) \left ( 1 - \sqrt{ \frac{v}{1 + v}} \right) \left(1 + \sqrt{\frac{v}{1+v}} \right) \\
& = (1+v) \left(1 - \frac{v}{1 + v}\right) \\
& = 1,
\end{align*}
and the last equivalence follows from $ \alpha = (1 - \sqrt{1 - \alpha})(1+\sqrt{1-\alpha})$. Thus, under the stationary condition, the necessary condition in \cref{eq:tail-condition} is equivalent to the sufficient condition in \cref{eq:gaussian_coefficient}.

We next determine the exact parameter range in which these two conditions can hold. The polynomial $q (2-q)(1+q)^2 = 2q + 3q^2 - q^4$ on $[0,1]$ has a derivative of
\begin{align}
\frac {d} {dq} \left( q (2-q) (1+q)^ 2 \right) & = 2 + 6q - 4q ^3 \nonumber \\
& \ge 2 + 2q \nonumber \\
& > 0, \qquad q \in [0 ,1].
\label{eq:thm1_eq3}
\end{align}

Its endpoint values are $0$ and $4$, respectively. Therefore, suppose $\sigma > \frac{1}{2}$, the value $\frac{1}{\sigma ^2}$ lies strictly between $[0, 4]$, so the intermediate value theorem and \cref{eq:thm1_eq3} give a unique $q_\sigma \in (0, 1)$ satisfying \cref{eq:critical_q}. Since $0< \alpha <1$, we have $\sqrt{1 - \alpha} \in (0, 1)$. The strict monotonicity just established and \cref{eq:tail-equivalent} imply
\begin{equation}
\sigma ^2\sqrt{1 - \alpha} (2 - \sqrt{1-\alpha} ) (1 + \sqrt{1  - \alpha} ) ^2 \le 1 \iff \alpha \ge 1 - q_\sigma ^2.
\label{eq:thm1_eq4}
\end{equation}

Suppose $0 < \sigma \le \frac{1}{2}$, then for every $\alpha \in (0, 1)$ we have
\begin{equation*}
\sigma^2 \sqrt{1 - \alpha} (2 - \sqrt{1 -\alpha}) (1 + \sqrt{1 - \alpha}) ^2 < 4 \sigma^2 \le1.
\end{equation*}

Thus the necessary tail inequality automatically holds in this parameter range.

Also, the stationary equation in \cref{eq:alpha_condition} may be re-written as
\begin{equation}
k^2 = \frac{ \alpha}{(1 - \alpha)(1 + \sigma^2 \alpha ^2) ^2}, \qquad 0 < \alpha < 1.
\label{eq:thm1_eq5}
\end{equation}

We now show that its right-hand side is strictly
increasing on the interval permitted by the tail inequality.
Direct differentiation gives
\begin{equation}
\frac{d} {d \alpha} \left( \frac{ \alpha}{(1 - \alpha) (1 + \sigma^2 \alpha^2) ^2} \right) = \frac{1 + \sigma^2 \alpha^2 (4 \alpha - 3)} {(1 - \alpha) ^2 (1 + \sigma^2 \alpha^2) ^3}.
\label{eq:thm1_eq6}
\end{equation}

To determine its sign, first suppose $\sigma ^2 \le 4$. For $0 \le \alpha \le \frac{3}{4}$,
\begin{align*}
\frac {d}{d \alpha} \left(\alpha^2(3 - 4 \alpha) \right) & = 6 \alpha - 12 \alpha ^2 \\
& = 6 \alpha ( 1 - 2\alpha).
\end{align*}

Therefore $\alpha ^2 (3 - 4 \alpha)$ attains its maximum $\frac{1}{4}$ at $\alpha = \frac{1}{2}$. For $\frac{3}{4} \le \alpha < 1$, the same expression is non-positive. Hence, throughout $(0, 1)$,
\begin{equation*}
1 + \sigma^2 \alpha^2 (4 \alpha - 3) \ge 1 - \frac{\sigma ^2}{4} \ge 0.
\end{equation*}

Equality is possible only when $\sigma^2 = 4$ and $\alpha = \frac12$. Thus the derivative in \cref{eq:thm1_eq6} is non-negative and vanishes at most at one point. The right-hand side of \cref{eq:thm1_eq5} is strictly increasing on $(0,1)$.

Suppose instead that $\sigma^2>4$. For $0 < q < 1$, expanding and factoring gives
\begin{align*}
q( 2 - q)(1 + q)^2 - 4q^2 & = 2q - q^2 - q^4 \\
& = q(1 - q)(q^2 + q + 2)\\
& > 0.
\end{align*}

Applying this at $q=q_\sigma$ and using \cref{eq:critical_q}, we obtain
\begin{align*}
4q_ \sigma ^2 & < q_\sigma (2 - q_\sigma)(1 + q_\sigma) ^2 \\ 
& = \frac{1}{\sigma^ 2},
\end{align*}
and consequently,
\begin{align}
1 - q_ \sigma ^2 & > 1 - \frac{1}{4 \sigma^2} \nonumber \\
& > \frac{15}{16} \nonumber \\
& > \frac34.
\label{eq:thm1_eq7}
\end{align}

Consequently, for every $\alpha \in [1-q_ \sigma^2, 1)$, we have $4 \alpha - 3 > 0$. Thus the numerator of \cref{eq:thm1_eq6} is strictly positive, and the right-hand side of \cref{eq:thm1_eq5} is strictly increasing on this interval.

For every $\sigma>\frac12$, the right-hand side of \cref{eq:thm1_eq5} therefore maps $[1 - q_\sigma^2, 1)$ continuously and strictly increasingly onto the interval beginning at its left-endpoint value and extending to $ +\infty$. We compute that endpoint from \cref{eq:critical_q}
\begin{align*}
\sigma^2(1 - q_\sigma^2)^2 & = \frac{(1-q_\sigma^2)^2} {q_ \sigma(2 - q_\sigma)(1 + q_\sigma)^2} \\
& = \frac{(1 - q_\sigma) ^2}{q_\sigma(2 - q_\sigma)},
\end{align*}
and
\begin{align}
1 + \sigma^2(1 - q_\sigma ^2) ^2 & = \frac{q_ \sigma(2 - q_\sigma) + (1 - q_ \sigma) ^2} {q_\sigma ( 2 - q_\sigma)} \nonumber \\
& = \frac{1}{q_ \sigma (2 - q_\sigma)}.
\label{eq:thm1_eq8}
\end{align}

Hence, substituting $\alpha = 1 - q_\sigma^2$ into \cref{eq:thm1_eq5} gives
\begin{align}
\frac{1 - q_\sigma^2} {q_\sigma^2 (1 + \sigma^2 (1 - q_\sigma ^2) ^2) ^2}
& = (1 - q_ \sigma^2)(2 - q_\sigma) ^2 \nonumber \\
& = k_c( \sigma) ^2.
\label{eq:thm1_eq9}
\end{align}

We then divide the parameters into two intervals. For $ \sigma \in (0, \frac{1}{2}]$, the stationary condition has a unique solution $\alpha \in (0,1)$ for any $k > 0$, and the tail inequality holds automatically. For $\sigma > \frac{1}{2}$, the tail inequality is equivalent to $\alpha \ge 1 - q_ \sigma ^2$. The stationary condition is strictly increasing in $\alpha$ on $[1 - q_\sigma^2 , 1)$, with its right-hand side ranging from $k_c (\sigma) ^2$ to $+ \infty$. Therefore, the two conditions admit a unique common solution if and only if $k \ge k_c( \sigma)$.

We also verify uniqueness of the stationary root in the entire interval $(0, 1)$ when $k \ge k_c(\sigma)$. For $\sigma \le 2$, the stationary equation is strictly increasing in $\alpha$. For $\sigma > 2$, the critical equation gives $q_\sigma < \frac{1}{4}$, and hence $k_c(\sigma)^2 > 1$. Moreover, for $\alpha \in (0 ,\frac34)$, we have
\begin{equation*}
\frac{\alpha}{(1 - \alpha)(1 + \sigma^2\alpha^2) ^2} \le \frac{48}{169} <  k^2.
\end{equation*}

Thus no stationary root lies in $(0, \frac34]$. Since the stationary equation is strictly increasing for $\alpha > \frac34$, the root is unique.

Suppose $k \in ( 0, k_c(\sigma))$. If a canonical globally optimal encoder $f^*$ has a finite positive-tail gain $\alpha$, Lemma~\ref{lem_4} would imply that $\alpha \in(0, 1)$ satisfies both the stationary condition and the tail inequality. This contradicts the parameter classification above. Since Lemma~\ref{lem_2} bounds $\frac{f ^*(x)}x$ strictly away from zero and infinity on the positive tail, its $\liminf$ and $\limsup$ must be distinct, which concludes \cref{eq:nonconvergent_gain}.

Suppose $k \in [ k_c(\sigma) , + \infty)$. The preceding classification gives a unique $\alpha \in(0, 1)$ satisfying both conditions. By their equivalence to the Gaussian coefficient condition, Lemma~\ref{lem_6} shows that $\mathcal N(0,\sigma^2\alpha^2)$ is the unique globally optimal output law, with the essentially unique linear controllers in \cref{eq:linear_optimal_pair}. The positive entropy remainder also guarantees uniqueness at $k=k_c(\sigma)$.

Finally, by symmetry of the Gaussian source and noise, the reflected controller pair
\begin{equation*}
x \to -f^*(-x), \qquad y \to -g^*(-y)
\end{equation*}
is also globally optimal, and its encoder is non-decreasing. Although reflection may change the continuity convention at jump points, two non-decreasing functions agreeing almost everywhere have the same finite asymptotic gain whenever either limit exists. Indeed, by choosing agreement points in $(x - 1, x)$ and $(x, x + 1)$, monotonicity bounds the value at $x$ between the two neighboring values, whose normalized limits coincide. Since
\begin{equation*}
\frac{-f ^*(-x)}{x } = \frac{f ^*(-x)}{-x},
\end{equation*}
a finite negative-tail gain would imply a finite positive-tail gain for a canonical globally optimal encoder, contradicting the non-tame result. The same $\liminf$--$\limsup$ separation therefore holds at $-\infty$. In the tame region, both tails are linear by \cref{eq:linear_optimal_pair}.
\end{proof}

The preceding results have implications far beyond asymptotic linearity, revealing a fundamental obstruction to the exact representation of globally optimal controllers within a much broader class of functions. We first recall the notion of $o$-minimality in \cite{10.1215/S0012-7094-96-08416-1} and adopt the following terminology.

\begin{definition}[$o$-minimality and tameness]
\label{def:tameness}

An expansion $\mathfrak R$ of the ordered real field  is called $o$-minimal if every subset of $\mathbb R$ definable in $\mathfrak R$, with real parameters, is a finite union of points and intervals. A function from $\mathbb R$ to $\mathbb R$ is called tame if its graph is definable, with real parameters in some $o$-minimal expansion of the ordered real field. It is called non-tame if no such expansion exists.

\end{definition}

We then show the optimal nonlinear encoder can not be defined in any $o$-minimal expansion.

\begin{corollary}[indefinable in any $o$-minimal expansion]
\label{cor_1}

For any parameters $\sigma \in (\frac12, \infty)$ and $0 < k < k_c(\sigma)$, the optimal encoder $f^ *$ of every globally optimal controller pair $(f^* , g^*)$ is not definable in any $o$-minimal expansion of the ordered real field.

\end{corollary}

\begin{proof}

In \cite{Wu2011}, the optimal encoder $f ^*$ is shown to admit a strictly increasing representative. Suppose a globally optimal encoder $f^*$ is definable in some $o$-minimal expansion of the ordered real field, then $\frac{f^*(x)}x$ is also definable and hence eventually monotone. By Lemma~\ref{lem_2}, this ratio is bounded almost everywhere on the positive tail and therefore converges to a finite limit.

Since $f^*$ agrees almost everywhere with its canonical non-decreasing representative, monotonicity transfers this limit to the canonical encoder by squeezing its value between agreement points in $(x-1, x)$ and $(x, x+1)$, which contradicts Theorem~\ref{thm_1}.

\end{proof}

This completes our characterization of linear optimality and tame representability over the positive parameter space. In the following section, we summarize our results and discuss its implications.

\section{Conclusions}

This paper reveals a fundamental structural distinction among globally optimal controllers in Witsenhausen's counterexample. We separate two regimes with fundamentally different properties. In the first regime, the globally optimal controller pair is essentially unique and linear, admitting an explicit representation that also enables exact evaluation of the optimal cost. In contrast, in the second regime, no globally optimal encoder is definable in any $o$-minimal expansion of the ordered real field, revealing a much deeper structural complexity beyond conventional nonlinear control.

The authors conjecture that this inherent complexity presents a substantial obstacle to constructing exact optimal controllers and establishing matching global bounds for the optimal cost. Consequently, numerical approximation may remain the most practical approach in the non-tame regime.

\section*{Acknowledgments}

The authors used AI tools to assist in preparing this manuscript. In particular, AI model helped expand the authors' existing mathematical arguments and preliminary manuscript into detailed derivations, and assisted with the organization and refinement of the exposition. The authors apologize for any non-accessible writing and will release a more polished version of this manuscript. Feel free to contact us if you find any issues.


\appendix
\input{numerical_validation}

\bibliographystyle{unsrt}  
\bibliography{references}  

\end{document}

%% file: numerical_validation.tex
\section{Numerical validation}
\label{app:numerical-validation}

\begin{figure}[!htbp]
\centering
\resizebox{0.98\linewidth}{!}{\input{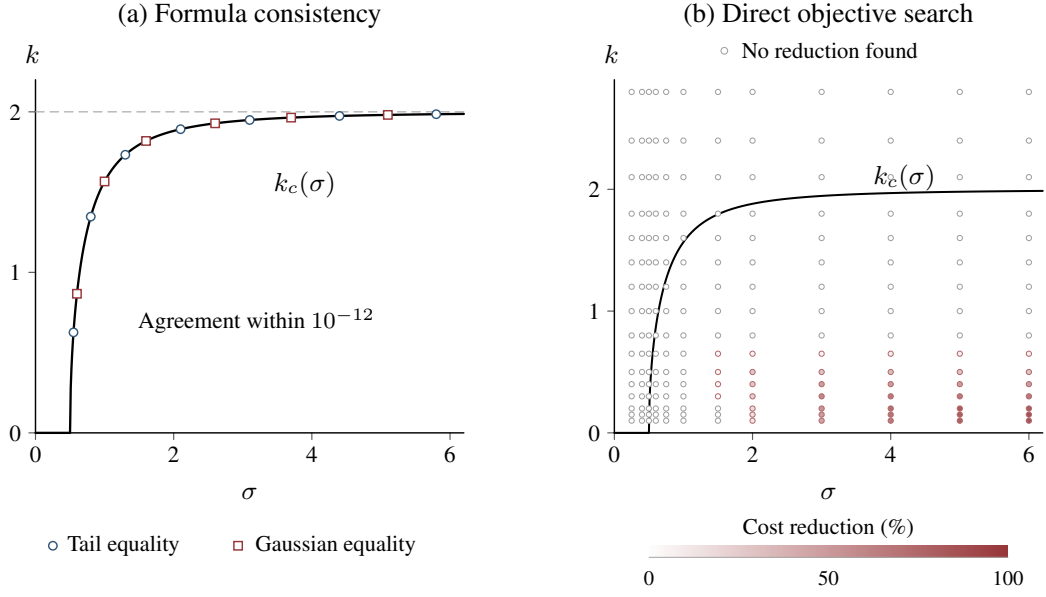}}
\caption{Numerical boundary tests. In (a), the curve is \cref{eq:intro-boundary}, and the symbols solve the stationary equation with equality in the tail condition or the Gaussian optimality condition. In (b), each circle is one independently searched parameter pair. Red shading gives the validated cost reduction, and gray circles indicate no reduction above the stated tolerance. The black curve is the theoretical boundary.}
\label{fig:boundary-validation}
\end{figure}

We test the boundary and the linear controller characterization through
formula comparisons and independent minimization of the original cost.
First, we compute the threshold from \cref{eq:intro-boundary} at $555$
source scales in $(1/2,6]$, including values within $10^{-8}$ of $1/2$.
We also solve the stationary equation together with equality in either
\cref{eq:tail-condition} or \cref{eq:gaussian_coefficient}.
The three thresholds agree within $10^{-12}$ at every tested point.
Figure~\ref{fig:boundary-validation}(a) illustrates this formula consistency.

For an independent test, we search \cref{eq:cost} at $192$ parameter
pairs formed by $12$ source scales in $[0.25,6]$ and $16$ weights in
$[0.1,2.8]$. The optimization uses only the objective and the Gaussian
laws. We optimize $81$ output levels on source cells of equal Gaussian
probability from eight symmetric starts and three unrestricted refinements,
and $17$ free values of a piecewise affine
policy on $[-5\sigma,5\sigma]$ from three starts, with affine tails.
The decoder is the conditional mean for each candidate. Gaussian cell
moments and affine segment convolutions are evaluated analytically
before observation quadrature.
The best linear policy is retained as a feasible candidate.
Every reported improvement is checked by independent adaptive quadrature,
with cost discrepancies below $10^{-12}$.

Figure~\ref{fig:boundary-validation}(b) shows the percentage cost reduction
relative to the best linear policy. An improvement is recorded only if
its absolute value exceeds $10^{-7}$ and ten times the error indicator
from quadrature refinement and independent integration. We find $39$ improvements among the $106$ tested points below
the boundary and none among the $86$ above it. Gray circles include $67$
points below the boundary where the finite searches found no improvement.
Such searches may miss gains confined to rare source tails.
At the classical point $(\sigma,k)=(5,0.2)$, the linear cost is $0.96$,
Witsenhausen's two level policy~\cite[Section~6]{counterexample} has
recomputed cost $0.40425320$, and the searched policy attains $0.17356729$.

\begin{figure}[!htbp]
\centering
\resizebox{0.98\linewidth}{!}{\input{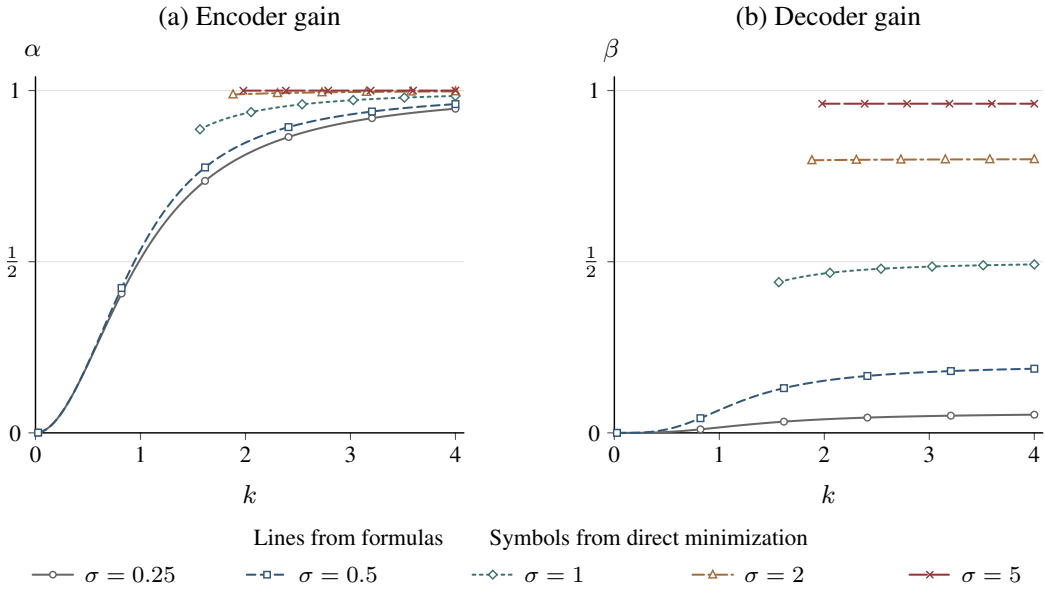}}
\caption{Validation of both linear controller gains. Lines show $\alpha$ and $\beta=\alpha^2\sigma^2/(1+\alpha^2\sigma^2)$ from \cref{eq:intro-linear-pair} and~\cref{eq:intro-linear-gain}. Symbols are independent minima of \cref{eq:numerical-linear-cost}. Each curve starts at $k_c(\sigma)$ when this threshold is positive. The numerical minimization uses objective values without substituting either gain formula.}
\label{fig:linear-validation}
\end{figure}

We separately test both gains in the linear policy pair $f(x)=ax$ and
$g(y)=by$ by minimizing its cost
\begin{equation}
\label{eq:numerical-linear-cost}
J_{\rm lin}(a,b)=k^2\sigma^2(a-1)^2+
\sigma^2a^2(1-b)^2+b^2.
\end{equation}
A scan of $4097$ values of $a\in[0,1]$ locates candidate minima,
including endpoint brackets, which are refined by golden section search.
For each $a$, the minimizing $b$ is recovered from three objective values.
A separate golden section search checks $b$ at the selected $a$.
Using $60$ decimal digits, we test $5742$ parameter pairs on or above
the boundary at $96$ source scales in $[0.1,20]$, including $67$ positive
boundary points and $84$ cases with $\sigma=1/2$.
Both gains and their costs agree with \cref{eq:intro-linear-pair}
and~\cref{eq:intro-linear-gain} within $10^{-12}$.
Repeating $141$ cases with $8193$ scan points preserves this agreement.
Figure~\ref{fig:linear-validation} shows representative gain curves.

\clearpage